\documentclass{aa}  
\usepackage{natbib}
\bibpunct{(}{)}{;}{a}{}{,} 
\usepackage{graphicx}
\usepackage{txfonts}
\usepackage{lipsum}
\usepackage{lineno}
\usepackage{booktabs}
\usepackage{subcaption}         
                          
\usepackage{lscape}             
                                
\usepackage{placeins}           
                                                   
\usepackage{hyperref}
\hypersetup{%
  colorlinks=true,
  linkcolor=blue,
  urlcolor=blue,
  citecolor=blue,
  bookmarksnumbered=true,
  bookmarksopenlevel=1,
  }

\begin{document}
\newcommand{\orcid}[1]{%
    \href{https://orcid.org/#1}{
        {%
            \includegraphics[width=0.3cm]{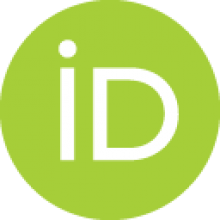}
        }{%
        }%
    }%
}
   \title{Stationary quasi-periodic pulsations in 20-second cadence TESS flares}

   \author{Aadish Joshi \inst{1,2} \orcid{0009-0007-1278-2600}
    \and Tom Van Doorsselaere \inst{1} \orcid{0000-0001-9628-4113}
    \and Daye Lim \inst{1,3} \orcid{0000-0001-9914-9080}
    \and Dario J. Fritzewski \inst{2} \orcid{0000-0002-2275-3877} 
        }

   \institute{Centre for mathematical Plasma Astrophysics (CmPA), 
   Department of Mathematics, KU Leuven, Celestijnenlaan 200B, 3001 Leuven, Belgium\\
             \email{tom.vandoorsselaere@kuleuven.be}
             \and
             Institute of Astronomy (IvS), KU Leuven, Celestijnenlaan 200D, 3001 Leuven, Belgium
             \and
             Solar-Terrestrial Centre of Excellence – SIDC, Royal Observatory of Belgium, Ringlaan -3- Av. Circulaire, 1180 Brussels, Belgium}

   \date{\today}

  \abstract 
   {Quasi-periodic pulsations (QPPs) are an inherent feature of solar and stellar flares. However, the mechanism behind them is debated hence it is necessary to further study them to obtain a complete picture of flares and their contribution to coronal heating.}
   {We analyze 20-second cadence TESS light curves from sectors 27 to 80 to detect stellar flares and QPPs.} 
   {Stellar flare detection was carried out using an automated detection routine based on autoregressive integrated moving average models. QPPs were detected using a Fourier model comparison test (AFINO).}
   {We detected 3878 flares across 1285 flaring stars. Notably, 61.2$\%$ of flares had a duration of less than 10 min. 61 QPPs were detected across 57 stars significantly expanding the current stellar QPP catalog. The detected periods of the QPPs were in the range of 42 to 193 seconds. In the diagram showing QPP periods against the flare duration a branch emerges. It shows a positive correlation with the flare duration, meaning longer duration flares host longer period QPPs.}
   {Our study detected short-period and sub-minute QPPs in stellar flares that have rarely been explored in other works. We find similar scaling laws between solar and stellar QPPs which indicates that QPPs in stellar flares might be analogous to the ones in solar flares as both show evidence of scaling with flare duration.}


   \maketitle

\nolinenumbers    
\section{Introduction}
\label{sec:introduction}
Solar flares are known to host quasi-periodic pulsations (QPPs) in both thermal and non-thermal emission on time scales from seconds to minutes with an occurrence rate of >80$\%$ \citep{2020STP.....6a...3K}. Currently, two mechanisms are proposed that could explain QPPs \citep[see][]{2016SoPh..291.3143V, 2018SSRv..214...45M, 2021SSRv..217...66Z}. Firstly, magnetohydrodynamic (MHD) wave modes within coronal loops, such as fast or slow magnetoacoustic waves are proposed to modulate flare emission by fluctuating plasma parameters (i.e density and magnetic field) in a recurring manner, leading to the observed pulsations \citep{1983SoPh...88..179E}. Secondly, reconnection itself can be inherently periodic, with instabilities and wave-like outflows naturally creating repetitive particle acceleration episodes and associated emissions \citep{2023ApJ...943..131K, 2009A&A...494..329M}. 

QPPs are not exclusive to the Sun but have also been observed in stellar flares, thereby raising the fundamental question of whether a similar mechanism is responsible for both solar and stellar QPPs. The first QPPs in a stellar flare were observed by \citet{1974A&A....32..337R} and several studies have since observed stellar QPPs in optical \citep{2003A&A...403.1101M}, ultraviolet \citep{2006A&A...458..921W}, microwave \citep{2004AstL...30..319Z} and X-ray \citep{2009ApJ...697L.153P}. Since the launch of the Kepler \citep{2010Sci...327..977B} and Transiting Exoplanet Survey Satellite (TESS) missions \citep{2014SPIE.9143E..20R}, it has been possible to systematically study stellar flares in the optical regime including the QPPs in their light curves. To this end, several key studies have found that QPPs on other stars have an occurrence rate in the range of 1.5$\%$ to 7$\%$ \citep{2015MNRAS.450..956B,2016MNRAS.459.3659P, 2022ApJ...926..204H, 2024ApJS..274...31B}. Most of the QPPs that have been detected through these surveys have a period of tens of minutes, which is generally higher than what we observe in solar flare QPPs. The lack of stellar QPPs in the period range of a few seconds to a few minutes could be attributed to the lack of higher-cadence data available. However, it is also possible that the detection method employed by previous works may not have been ideal for detecting low periodicity in slower cadence observations when compared to solar flare observations which typically have observing cadences around a couple of seconds. In solar-flares, QPPs with sub-minute periods are commonly seen. Radio and X-ray observations of stellar flares have revealed QPPs with a sub-minute period range \citep[see for e.g.][]{1982ApJ...263L..79G, 1986ApJ...305..363L}. However, the studies focused on individual stars. Hence, to get a complete picture of stellar QPPs, it is imperative to also search for such low-period pulsations in optical QPPs as well and obtain a statistical catalog of such events. Using Kepler short cadence data, \citet{2015MNRAS.450..956B} found  QPPs with periods from 4.8 to 14 minutes. \citet{2016MNRAS.459.3659P} expanded the catalog of Kepler QPPs by identifying 56 flares showing evidence of QPPs with periods from 4.6 to 93 minutes. It was found that QPP periods do not correlate with the physical stellar parameters \citep{2016MNRAS.459.3659P}. With the 2-minute cadence TESS data, \citet{2021SoPh..296..162R} found eleven QPPs on seven individual stars with a period ranging from 10 to 71 minutes. \citet{2021csss.confE.272M} detected nine QPPs with periods between 2 and 8 minutes in 20-second TESS data. Recently, \citet{2022ApJ...926..204H} utilized the TESS 20-second cadence observations to identify 17 QPPs and most of the QPPs had periods between 3 to 25-minutes.

In this paper, we analyze the 20-second cadence TESS observations from sectors 27 to 80 and present the largest collection of QPPs studied in the TESS catalog to date. In order to detect flares in the TESS observations, we developed an automated flare detection routine, and for detecting the QPPs on the flares we use the Fourier power spectrum method as outlined in \citet{2015ApJ...798..108I, 2016ApJ...833..284I}. As noted earlier, the QPPs that were identified in other stellar QPP studies have periods in tens of minutes which is why this study mainly focuses on detecting sub-minute and short-period QPPs within 3-minutes. In this way, we provide a more comprehensive catalog of QPPs from seconds to minutes to complement other longer-period catalogs. 

The paper is divided as follows: In Section \ref{sec:obs} we briefly describe TESS observations. In Section \ref{sec:flaredetection}, we explain our flare detection method and the results from our flare detection analysis. In Section \ref{sec:qpps}, we discuss the QPP detection method along with the results from the QPP analysis. Lastly in Section \ref{sec:conclusion}, we summarize our findings and conclude. 
\section{Observations}
\label{sec:obs}
The Transiting Exoplanet Survey Satellite \citep{2014SPIE.9143E..20R} has been conducting an all-sky survey since 2018. At the time of writing this paper, TESS has completed surveying 80 sectors. TESS has four CCD cameras that provide a field of view of 24\textdegree x 96\textdegree. It divides both the northern and southern hemispheres of the sky into 13 sectors each, making a total of 26 sectors. Each sector is observed for 27.4 days \citep{2015JATIS...1a4003R}. The onboard cameras capture images every two seconds, and these images are subsequently combined into sets of 8 and 60, yielding effective cadences of 20 seconds and 120 seconds, respectively. All the TESS targets have a designated number, also called the TESS Input Catalog (TIC) which is used for target selection for the TESS mission.

Since TESS cycle-3 it has been possible to propose targets for observation with the 20-second cadence. To this end, our study selected all the stars observed in 20s cadence in Sector 27-80. The data was bulk downloaded using the cURL scripts from the Mikulski Archive for Space Telescopes \footnote{https://archive.stsci.edu/tess/bulk\_downloads.html}. We analyzed a total of 66,527 unique objects. However, because the targets in the 20-second catalog are specifically selected targets in the Guest Investigator program, our detection rates are inherently biased. Consequently, the measured distributions may not accurately reflect the entire stellar population. 
\section{Lightcurve analysis}
\label{sec:flaredetection}
\subsection{Detection method}
Stellar light curves obtained from TESS often exhibit long-term trends and temporal correlations due to instrumental effects, stellar rotation, or other astrophysical phenomena. To effectively isolate and identify flares in the light curves of our sample stars, we apply a time-series modeling technique to achieve a stationary time-series. Our goal is to remove long-term correlations and enhance the detection of transient events by analyzing the residuals.
\subsubsection{ARIMA}
ARIMA is a statistical modeling technique that is commonly used for time-series forecasting. It combines three core components: autoregression (AR), differencing to achieve stationarity (I for "integrated"), and a moving average model (MA) \citep{box1970time}. The model can be written as:\\
\begin{equation}
\label{arimaeq}
  y'_{t} = c + \phi_{1}y'_{t-1} + \cdots + \phi_{p}y'_{t-p}
     + \theta_{1}\varepsilon_{t-1} + \cdots + \theta_{q}\varepsilon_{t-q} + \varepsilon_{t},
\end{equation}
where $y'_{t}$ is the differenced time-series, \(\phi_i\) for $i$ = 1, 2,.., $p$ are the autoregressive coefficients for lag \(i\), \(\epsilon_t\) is the white noise error term at time \(t\), with \(\epsilon_t \sim \mathcal{N}(0, \sigma^2)\) and \(\theta_j\) for $j$ = 1, 2,..., $q$ are the moving average coefficients for lag \(j\). In other words, the model can be defined by three components ARIMA($p$, $d$, $q$). Where $p$ (autoregressive order) is the number of lagged observations in the model. In Eq. \ref{arimaeq}, the terms $\phi_1 y'_{t-1},\ \phi_2 y'_{t-2},\ \dots,\ \phi_p y'_{t-p}$ —each multiplied by its corresponding coefficient correspond to the autoregressive (AR) component. $q$ denotes the order of the moving average components, which corresponds to the number of lagged forecast errors included. The terms $ \theta_1 \varepsilon_{t-1},\ \theta_2 \varepsilon_{t-2},\ \dots,\ \theta_q \varepsilon_{t-q}$ represent the moving average (MA) component, where each \(\theta_j\) is the coefficient corresponding to the error at lag \(j\). To make a non-stationary time-series into a stationary time-series we subtract the differences between the components (differencing). Given a timeseries $y_{t}$ we define its first-order differenced series as $y'_{t}$:
\begin{equation}
y'_{t} = y_{t} - y_{t -1}
\end{equation}
Here, $y'_{t}$ represents the change between consecutive observations, effectively removing linear trends. This first-order differencing corresponds to $d$ = 1  in the ARIMA model and is often sufficient to achieve stationarity for many time series, but if trends persist, higher-order differencing can be applied. 

To systematically identify a suitable ARIMA model, we utilized the auto.arima() function from the pmdarima library in Python \citep{pmdarima}\footnote{https://github.com/alkaline-ml/pmdarima}. This function conducts a search over a range of possible ($p, d, q$) parameters and selects the model with the lowest Akaike Information Criterion (AIC). By minimizing the AIC, the function attempts to find a model that provides a good balance between complexity and fit. We varied the parameters ($p,d,q$) within the ranges  $p \in [0,3]$, $d \in [0,2]$ and $q \in [0,3]$ to explore different model structures.
\begin{figure*}
\centering
  \includegraphics[width=18cm]{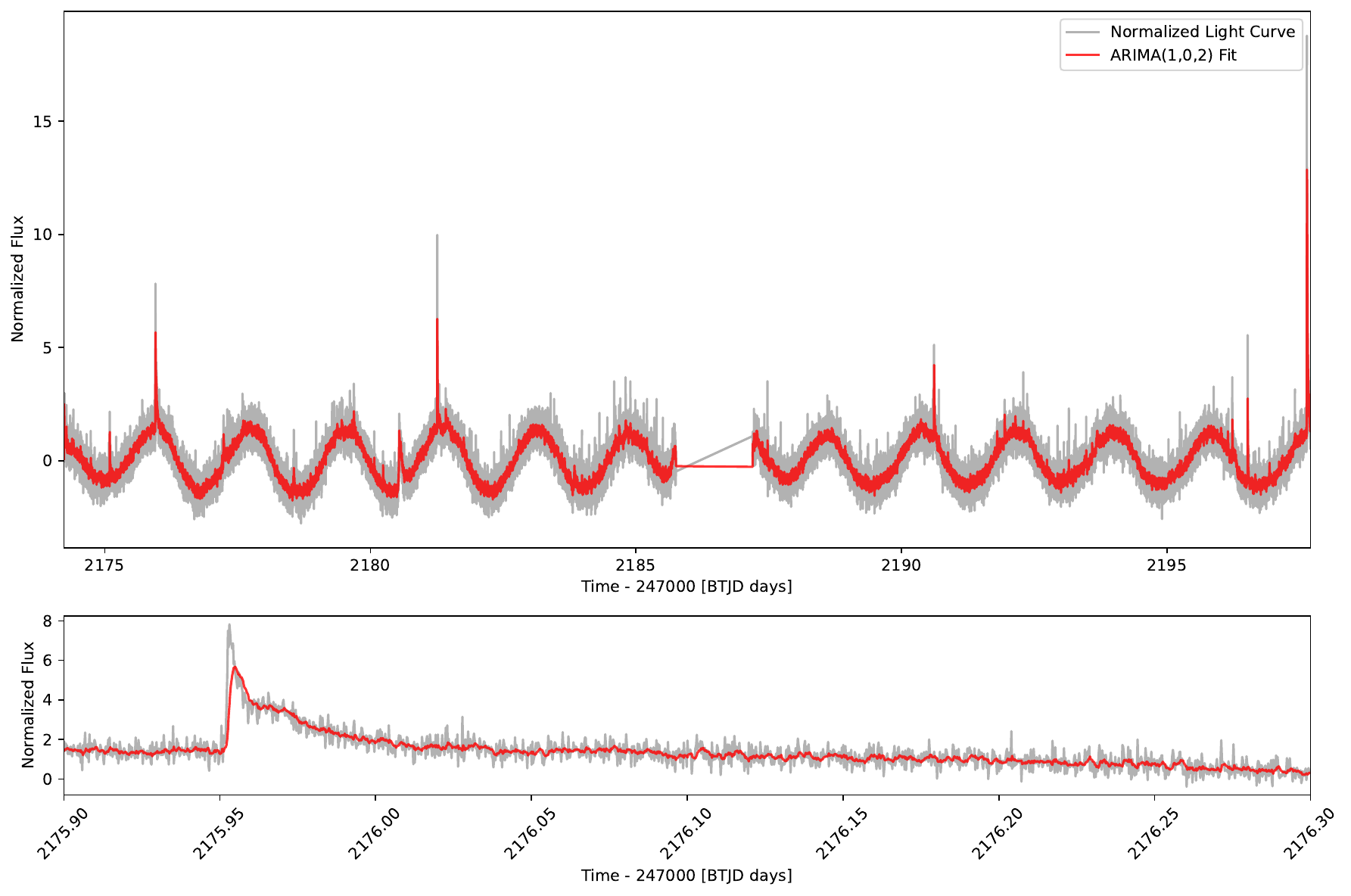}
  \caption{ARIMA model fit to the normalized light curve of TIC 1273249 observed in TESS Sector 32. The gray line represents the observed light curve, characterized by significant variability and periodic flaring events, while the red line shows the ARIMA model's predicted trend. The lower panel shows a zoomed view of a subset of the lightcurve.}
  \label{fig:residuals}
\end{figure*}
These parameter ranges were selected to balance simplicity and performance, ensuring the models are sufficiently flexible to capture the underlying patterns in the data while minimizing the risk of overfitting. We further explored restricting $p$ to the interval $[1,3]$ to ensure the model always includes an auto-regressive component. However, this constraint did not alter the flare detection outcomes. Fig. \ref{fig:residuals} illustrates the ARIMA model fit for TIC 1273249. The ARIMA(1,0,2) model was selected based on the Akaike Information Criterion (AIC), which was lower for this model. 

Once the best ARIMA model is selected, we generate predictions for the entire normalized flux time-series. By subtracting these predictions from the actual normalized flux, we obtain a residual series. Ideally, if the ARIMA model captures all underlying structure except flares, the residuals should resemble white noise, with a mean near zero and no significant autocorrelation.
\begin{figure*}[h!]
\centering
\includegraphics[width=\textwidth]{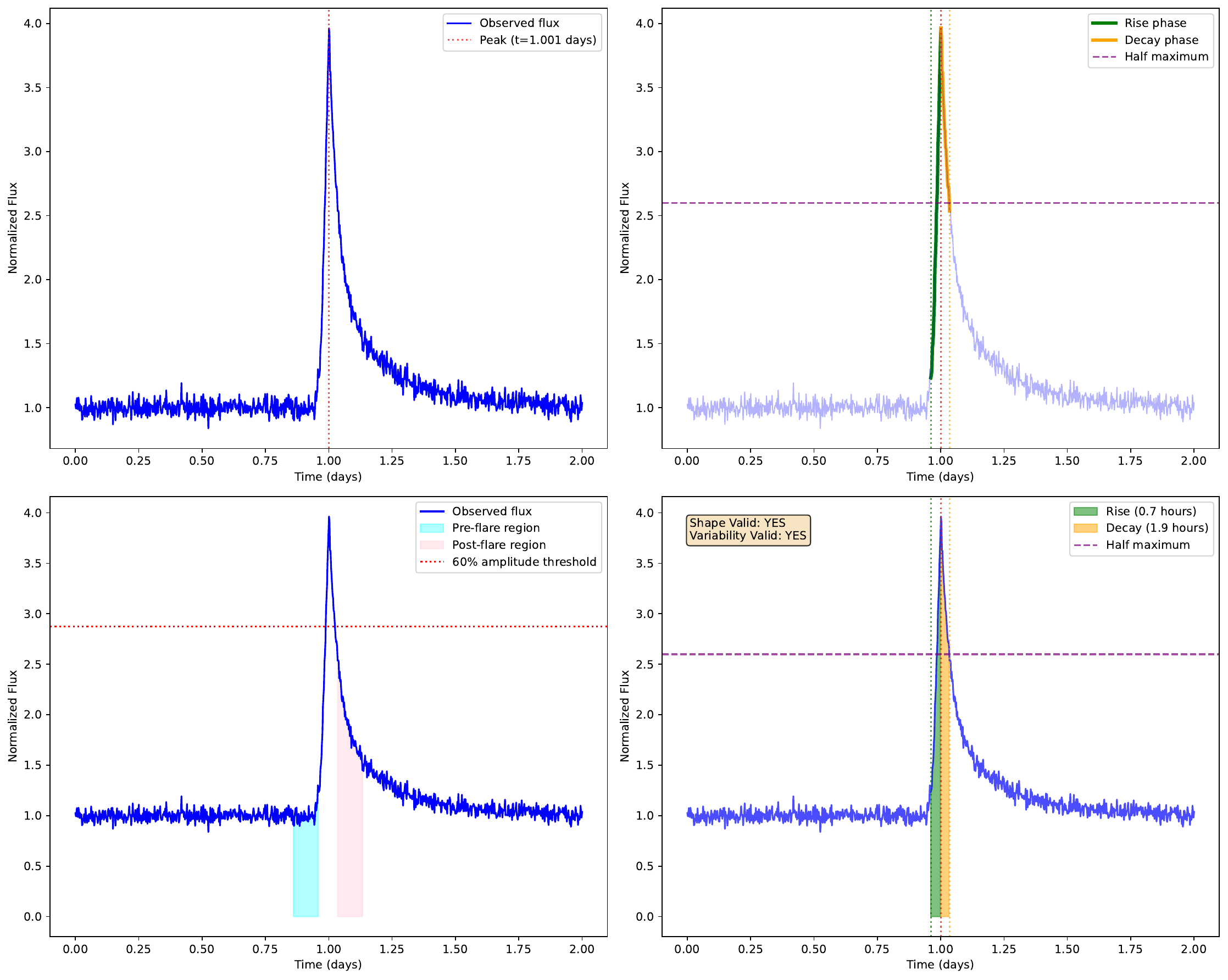}
\caption{Top left: A simulated flare generated using the template from \citet{2014ApJ...797..122D}, with the flare peak marked by a red dotted line. Top right: Determination of the flare’s start and end times. Bottom left: Assessment of local variability with designated baseline regions—pre-flare data (cyan) and post-flare data (pink). The red dotted horizontal line indicates the 60\% amplitude threshold. Bottom right: Detailed examination of the flare shape, highlighting the rise phase (green) and decay phase (orange). Key parameters including rise time, decay time, alongside the outcomes of the shape and variability checks.}
\label{fig:detection_routine}
\end{figure*}
\subsubsection{Detection routine}
Once the residuals are computed, we flag the residuals that exceed three times the standard deviation. Potential flare events manifest as large, positive excursions in the residuals and last longer than a single data-point. After identifying candidate indices where residuals exceed the threshold, we cluster continuous or near-continuous points to form a flare “event". To qualify as a physically plausible flare candidate, an event must exhibit a minimum of eight consecutive data points. This threshold was determined empirically, ensuring that only genuine flare-like events are identified while effectively filtering out false positives. Due to this condition, we may miss shorter flare events. This criteria does not affect the detection of QPPs because to have QPPs a flare must be sufficiently long anyway. We should also note that the choice of ARIMA parameters can affect the number of elevated residual points above the 3$\sigma$ threshold. A higher $p$ and $q$ could reduce residual variance by capturing more patterns, potentially leading to fewer elevated residuals. If the model ends up using an incorrect value of differencing it can increase residual variance and the number of elevated points \footnote{Following the detection of candidate flares, all subsequent analyses including flare filtering are conducted using the original normalized flux rather than the flux predicted by the ARIMA model.}.

A physically realistic stellar flare generally exhibits a fast rise and exponential decay \citep{2024LRSP...21....1K}. To check, we verify that (i) the flux increases leading up to the flare peak and decreases afterward, (ii) the flare peak is well above the local baseline (the standard deviation in the 50 data points before and after the event must not exceed $60\%$ of the flare amplitude), and (iii) the event displays a pronounced asymmetry in rise versus decay times. These steps are illustrated in Fig. \ref{fig:detection_routine}.

We split the candidate into two segments: a rise phase (from start to peak) and a decay phase (from peak to end). The start time of the flare is the earliest time preceding the peak where the normalized flux begins a monotonic rise. On the other hand, the end time is defined as the flux decays below the half-maximum level. This approach to determining the flare's end time is adopted from the criteria used in Geostationary Operational Environmental Satellite (GOES) X-ray flare observations. By examining these segments, we verify that the flux increases before reaching the peak and subsequently decreases afterward by checking that the rise segment is higher than its initial flux, and that the final flux of the decay segment is lower than its initial flux.

Next, we ensure that the flare’s peak is genuinely significant with respect to the local baseline. We calculate a median flux level from a short interval of 50 data points which corresponds to approximately 16 minutes immediately preceding the event’s start. Similarly, we calculate the standard deviation of the flux in the regions before and after the event, each consisting of 50 data points which must be less than or equal to 60$\%$ of the flare's amplitude. If the peak does not stand out above the surrounding noise level, it is likely just a random fluctuation rather than a real astrophysical flare. 

Finally, we assesses the symmetry of the event. True stellar flares are characteristically asymmetric: they brighten rapidly and then cool off more gradually \citep{2014ApJ...797..122D}. To quantify this asymmetry, we measure the ratio of the event's rise time to its decay time. An event is considered too symmetrical if this ratio falls between 0.5 and 2.0. Specifically, if the symmetry ratio satisfies: \[
0.5 < \text{symmetry\_ratio} < 2.0
\] the event is rejected. This range indicates that the rise and decay times are too similar, and the event may not be a flare but could instead be a periodic signal or another symmetric artifact. By contrast, acceptable flare candidates must have a symmetry ratio less than or equal to 0.5 (indicating a rapid rise and slower decay) or greater than or equal to 2.0 (indicating significant asymmetry in timing). A similar approach is taken in \citet{2017ApJS..232...26V}. Both emphasize FRED-like profiles and significance thresholds to exclude noise. While \citet{2017ApJS..232...26V} use a double threshold on flux and its derivative, alongside exponential decay fitting, our method quantifies asymmetry through rise/decay time ratios and enforces strict local noise constraints. Both aim to filter instrumental artifacts and periodic signals but differ in specific metrics (e.g., symmetry ratio vs. prewhitening and polynomial detrending). 

Additionally, we also discard events that exceed a total duration of more than 4-hours as we detected certain false detections that were longer than this threshold. However, we should note that events greater than 4 hours have been detected in other stellar flare studies e.g, \citet{2013ApJ...773..156A} in which they analyzed a megaflare on YZ CMi lasted over 7 hours, which is confirmed as a genuine stellar flare with a well-documented FRED-like profile and oscillatory signatures. So it is important to note that the 4-hour cutoff risks rejecting rare but physically real stellar megaflares.
\begin{figure*}[h!]
\centering
\includegraphics[width=0.48\textwidth]{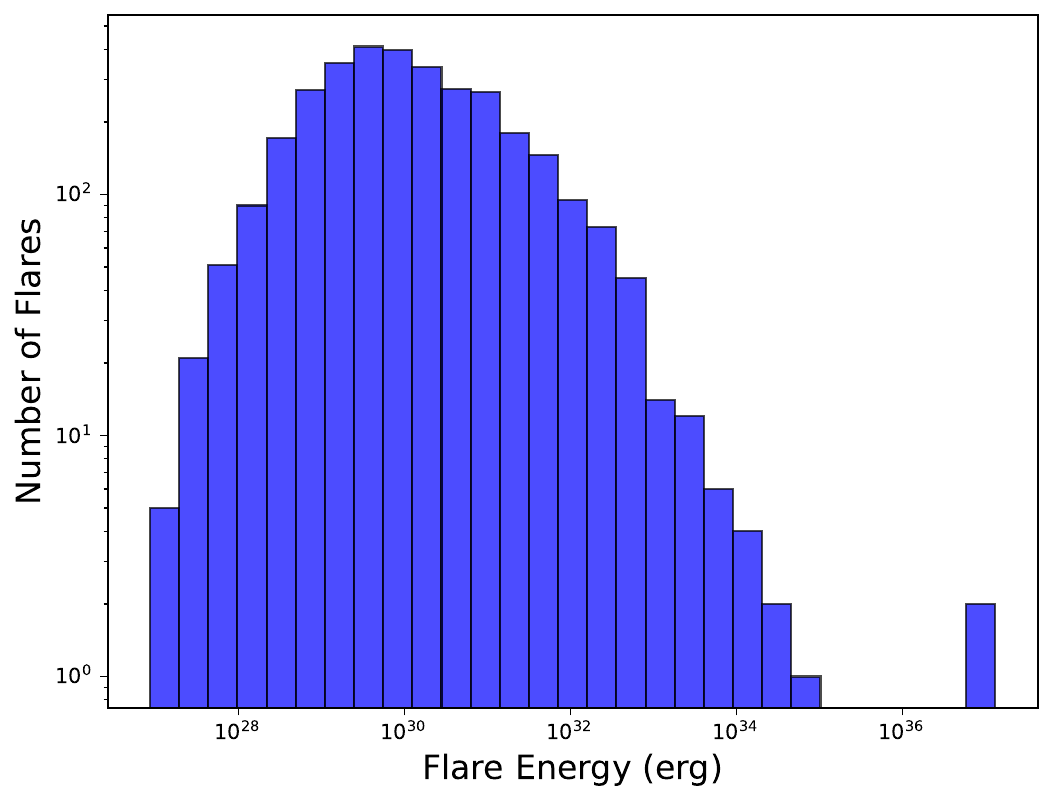}
\hfill
\includegraphics[width=0.48\textwidth]{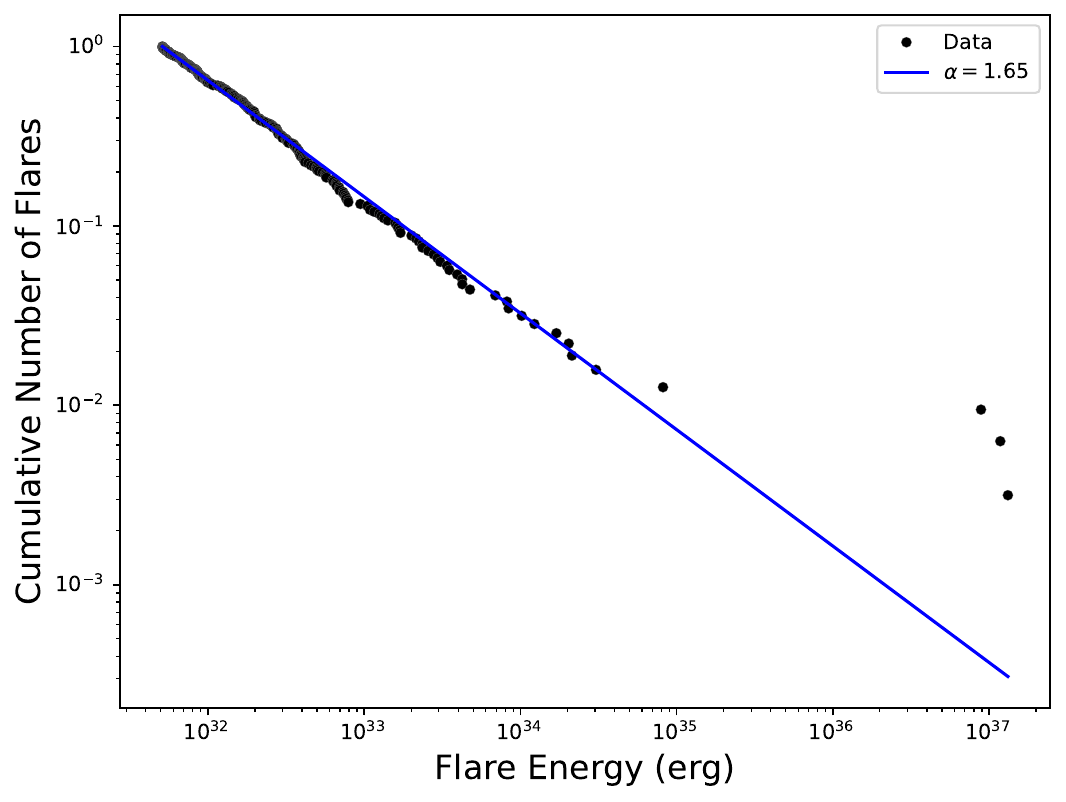}
\caption{Distribution of flare energy (left) and the cumulative flare frequency distribution with a power-law fit (right). The blue line represents the power-law fit.}
\label{fig:Energy_20s}
\end{figure*}
\begin{figure*}[h!]
\centering
\includegraphics[width=0.48\textwidth]{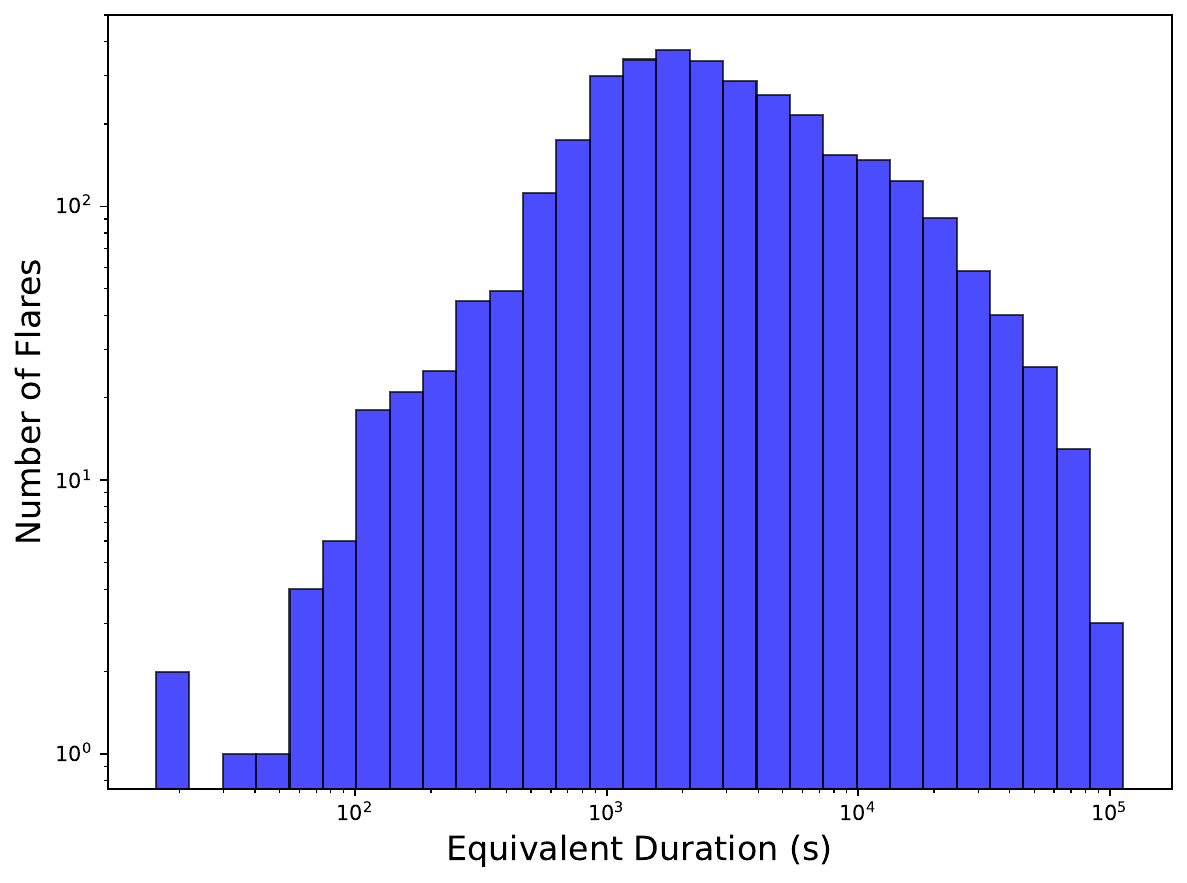}
\hfill
\includegraphics[width=0.48\textwidth]{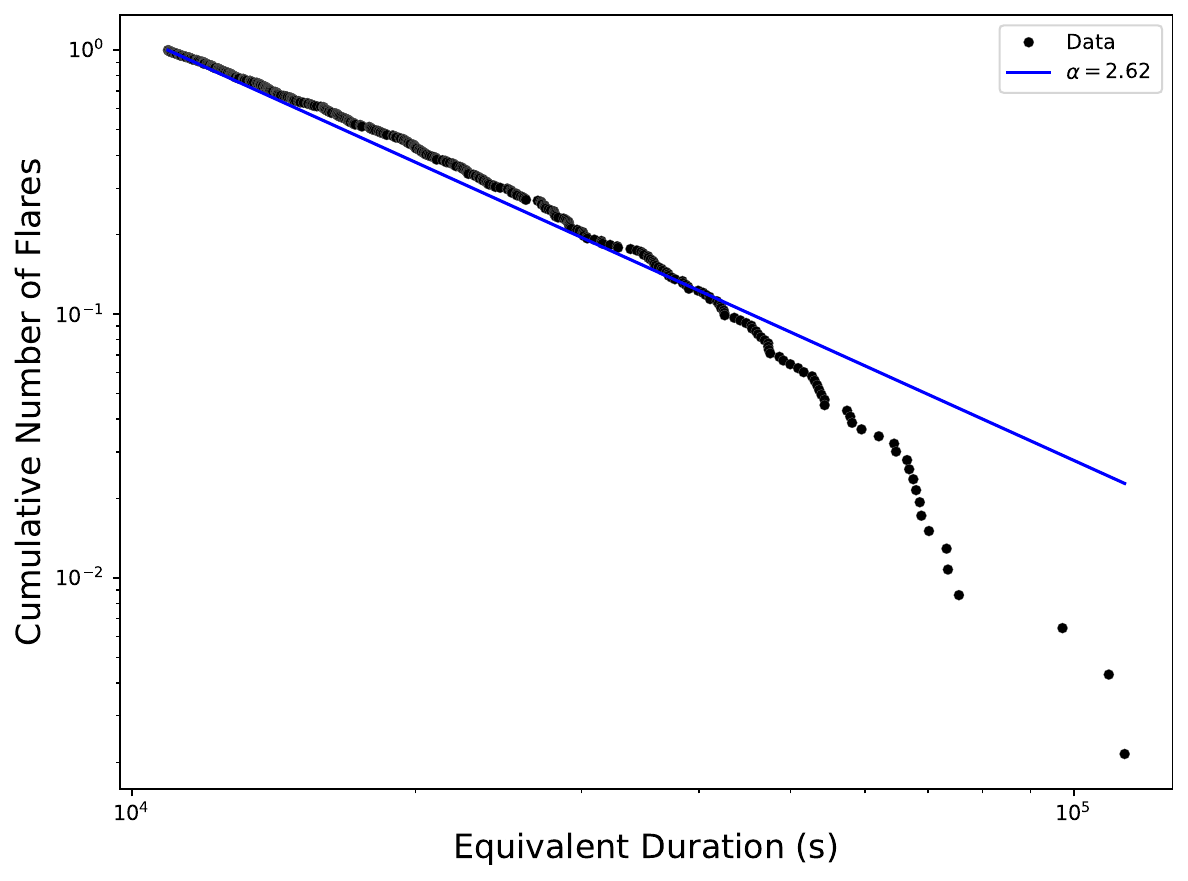}
\caption{Distribution of equivalent duration (left) and the cumulative frequency distribution with a power-law fit (right). The power-law fit is denoted using the blue line. Unlike the power-law exponent $\alpha$ found for flare energies, the exponent for equivalent duration (ED) is higher, as expected, since ED is not influenced by the instrument sensitivity of TESS.}
\label{fig:ED_20s}
\end{figure*}
\subsection{Flare properties}
Once a candidate flare passes all validation checks, we accept it as a confirmed flare and proceed to measure its key properties using the normalized flux rather than the ARIMA-predicted flux so that these measurements directly reflect the observed brightness changes. However it is important to note that because the normalized flux is derived from the original light curve (which is not detrended), any slow baseline variations or trends for instance, those induced by stellar rotation could shift the baseline level. This shift might lead to an overestimation or underestimation of the flare’s amplitude and, similarly, might contaminate the excess flux calculated for energy estimation if an upward trend is present during the flare event.

In practice, we first determine the flare amplitude by comparing the peak flux to the baseline level. We next measure the duration of the event, defined as the time from the start to the end of the flare, which is useful for classifying the energy release timescale. If stellar parameters and spectral information are available, we then estimate the total flare energy by integrating the flux excess over the flare’s duration. Finally, we measure the equivalent duration (ED), which represents how long the star would need to emit at its quiescent luminosity to match the total energy released during the flare, thereby providing a dimensionless measure of flare strength.

To compute the energy of a stellar flare, we follow the method outlined by \citet{2013ApJS..209....5S} and adopted by \citet{2020AJ....159...60G}. This approach models the flare's continuum emission as a blackbody spectrum at a temperature of \( T_{\text{flare}} = 9000\,\text{K} \) \citep{2013ApJS..207...15K, 1991ARA&A..29..275H,1992ApJS...81..885H}. We calculate the stellar luminosity \( L_{\star} \) using the Stefan-Boltzmann law:
\begin{equation}
    L_{\star} = 4\pi R_{\star}^2 \sigma_{\text{SB}} T_{\star}^4,
\end{equation}
where \( R_{\star} \) and \( T_{\star} \) are the stellar radius and effective temperature, respectively, and \( \sigma_{\text{SB}} \) is the Stefan-Boltzmann constant. We obtain \( R_{\star} \) and \( T_{\star} \) from the TESS Input Catalog (TIC). If \( R_{\star} \) is not available, we estimate it using empirical relations for main-sequence stars from \citet{2013ApJS..208....9P}. To account for the TESS instrument's sensitivity, we integrate the Planck function weighted by the TESS response function over the instrument's wavelength range, obtaining integrated fluxes \( F(T_{\star}) \) and \( F(T_{\text{flare}}) \) for the star and the flare, respectively. The ratio of these fluxes provides a scaling factor \( S \):
\begin{equation}
    S = \frac{F(T_{\star})}{F(T_{\text{flare}})}.
\end{equation}
Assuming the flare covers a small portion of the stellar surface, we estimate the flare area \( A_{\text{flare}}(t) \) from the relative flux increase \( (\Delta F / F)(t) \):
\begin{equation}
    A_{\text{flare}}(t) = \left( \frac{\Delta F}{F}(t) \right) \pi R_{\star}^2 S.
\end{equation}
The flare luminosity \( L_{\text{flare}}(t) \) is then:
\begin{equation}
    L_{\text{flare}}(t) = \sigma_{\text{SB}} T_{\text{flare}}^4 A_{\text{flare}}(t),
\end{equation}
and the total energy emitted by the flare \( E_{\text{flare}} \) is calculated by integrating \( L_{\text{flare}}(t) \) over the flare duration:
\begin{equation}
    E_{\text{flare}} = \int_{t_{\text{start}}}^{t_{\text{end}}} L_{\text{flare}}(t) \, dt.
\end{equation}
For detailed equations and computational steps, we refer the reader to \citet{2013ApJS..209....5S} and \citet{2020AJ....159...60G}.

In addition to estimating the total energy of a flare, we calculate the ED to provide an instrument-independent measure of the flare's strength. ED represents the time interval during which the star, at its quiescent luminosity, would emit the same amount of energy as was emitted during the flare \citep{2019A&A...622A.133I, 1972Ap&SS..19...75G}. ED is defined as:
\begin{equation}
\label{eqn}
\text{ED} = \int_{t_{\text{start}}}^{t_{\text{end}}} \left( \frac{F(t) - F_q}{F_q} \right) dt,
\end{equation}
where: \( F(t) \) is the observed flux at time \( t \), \( F_q \) is the quiescent flux of the star, \( t_{\text{start}} \) and \( t_{\text{end}} \) are the start and end times of the flare. Assuming that the flux is normalized such that \( F_q = 1 \), the equation simplifies to:
\begin{equation}
\label{eqn:ed_normalized}
\text{ED} = \int_{t_{\text{start}}}^{t_{\text{end}}} \left( F(t) - 1 \right) dt.
\end{equation}
In practice, observational data are discrete. Therefore, we approximate the integral as a sum over the observed time intervals:
\begin{equation}
\label{eqn:ed_discrete}
\text{ED} \approx \sum_{i=1}^{N} \left( F_i - 1 \right) \Delta t_i,
\end{equation}
where: \( F_i \) is the normalized flux at the \( i \)-th observation, \( \Delta t_i = t_{i+1} - t_i \) is the time interval between consecutive observations, \( N \) is the total number of observations during the flare. ED is independent of the star's distance and relies solely on relative flux measurements. 

\subsection{Determining rotation period}
Previous studies have indicated a strong correlation between stellar rotation rate and occurrence rate of flares \citep{2016ApJ...829...23D}. Hence, it is important to check if the rotation rate has any relation with the periods of the pulsations as well. For stars exhibiting QPPs in their flares, we obtained the stellar rotation rate from the catalog by \citet{2024AJ....167..189C}. However, not all stars were included in their catalog. We determined rotation periods for the remaining stars by computing a Lomb-Scargle periodogram with Lightkurve\footnote{https://github.com/lightkurve/lightkurve}. We searched periods between 0.5–14 days, typical for low-mass main-sequence stars, then adopted the highest-power peak as the best-fit rotation period. To confirm this peak represented the fundamental rotational signature rather than a harmonic, we folded each light curve at that period to check phase consistency and also examined the periodogram at twice that value. The rotation rate we identified on all the 57 stars ranges between 0.5d to 13d.

\subsection{Results from flare detection}
We found 3878 flares across 1285 flaring stars. The detected flare list is given in Table \ref{table:flare_table}. We note that the number of flares identified by our pipeline likely underestimates the total flaring activity in these stars, due to the pipeline’s reduced sensitivity to low-amplitude events as well as the likelihood of false negatives if the ARIMA model fits too closely to the flare profile, as the residuals might not show a sufficiently large deviation to cross the flare detection threshold, thereby causing genuine flares to be missed. For example, \citet{2020AJ....159...60G} carried out the first stellar flare study using TESS, analyzing the initial two sectors of 2-minute cadence data, which included approximately 8,000 flares from 1,228 flaring stars. They found a flare occurrence rate of 4.9$\%$. \citet{2022ApJ...935..143P} analyzed 39 TESS sectors of 2-minute cadence detecting over 140,000 individual flares on 25,229 stars. They highlighted that 7.7\% of the sample exhibited flaring behavior. Recently, \citet{2024A&A...689A.103Z} carried out flare detection for 20-second cadence light curves from 2020 to 2023 and detected 32,978 flare events associated with 5463 flaring stars. By comparison, our study yields a substantially lower occurrence rate of 1.6$\%$. Our requirement that the local standard deviation must be less than 60$\%$ of flare amplitude may exclude many valid events. Additionally, when pre-flare median flux is elevated by background variations, lower amplitude flares become difficult to distinguish. Furthermore, our classification criteria requiring at least 8 consecutive data points above the 3$\sigma$ threshold inherently prevents the detection of flares shorter than 160 seconds in duration. Nonetheless, this limitation does not hinder our scientific aim of detecting short-period QPPs.

\citet{2024A&A...689A.103Z} found 55$\%$ of all flare events in their 20-second cadence observations were less than 8-minutes. Likewise, in the low-mass regime, most flares detected by \citet{2022ApJ...926..204H} lasted under 10 mins. Consistent with their results, our analysis shows that 61.22$\%$ of the flares detected using 20-second cadence data had durations shorter than 10 minutes. This likely arises from high-cadence observations, which provide more data points and thus make short flares easier to detect. By contrast, the same short flares may blend into only one or two bins at lower cadence, producing a bias toward detecting longer flares at lower time resolution.
\begin{table*}
\centering
\caption{Flares detected by our pipeline with the various flare parameters.}
\setlength{\tabcolsep}{8pt} 
\resizebox{\textwidth}{!}{
\begin{tabular}{lcccccccc}
\hline\hline
\noalign{\smallskip}
TIC ID & $T_{\rm eff}$ (K) & Start time (TBJD) & End time (TBJD) & Peak time (TBJD) & Amplitude & Duration (days) & Flare energy (erg) & ED (s) \\
\noalign{\smallskip}
\hline
\noalign{\smallskip}
70111 & 3434 & 2344.458 & 2344.473 & 2344.462 & 0.25108 & 0.01412 & 9.30E+30 & 16898.84 \\
134947 & 3286 & 2350.105 & 2350.111 & 2350.107 & 0.383551 & 0.005555 & 6.63E+29 & 8177.108 \\
593230 & 3806 & 2194.986 & 2194.989 & 2194.986 & 0.034909 & 0.003009 & 6.64E+29 & 2283.266 \\
\noalign{\smallskip}
\hline
\multicolumn{9}{c}{\dots \dots \dots} \\
\hline
\noalign{\smallskip}
1400807589 & 5391 & 3384.017 & 3384.019 & 3384.018 & 0.008179 & 0.001852 & 1.71E+30 & 890.4324 \\
1400807589 & 5391 & 3393.721 & 3393.726 & 3393.722 & 0.008982 & 0.004861 & 7.91E+30 & 1450.681 \\
1400807589 & 5391 & 3393.913 & 3393.920 & 3393.915 & 0.01599 & 0.006713 & 2.70E+31 & 4013.727 \\
\noalign{\smallskip}
\hline
\end{tabular}
}
\tablefoot{Columns include the TIC ID, effective temperature ($T_{\rm eff}$) in Kelvin, start, end and peak times of each flare event, relative amplitude of the flare, duration in days, flare energy in erg and equivalent duration (ED) in seconds. Note:  Flare energy is not calculated for entries lacking $T_{\rm eff}$ data. The full table can be found at the GitHub repository: https://github.com/aadishj19/QPPs-in-TESS-flares}
\label{table:flare_table}
\end{table*}
\begin{figure*}
\centering
\includegraphics[width=0.48\textwidth]{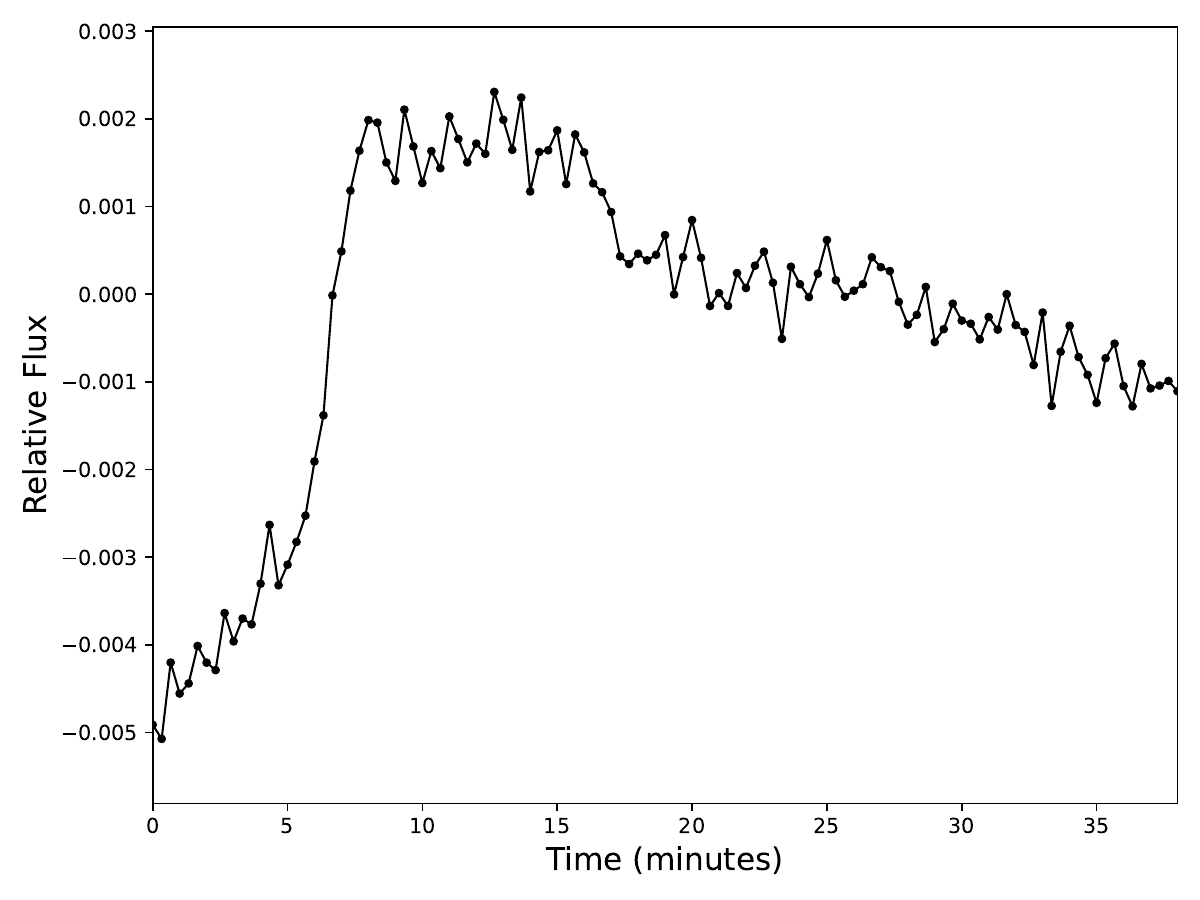}
\hfill
\includegraphics[width=0.48\textwidth]{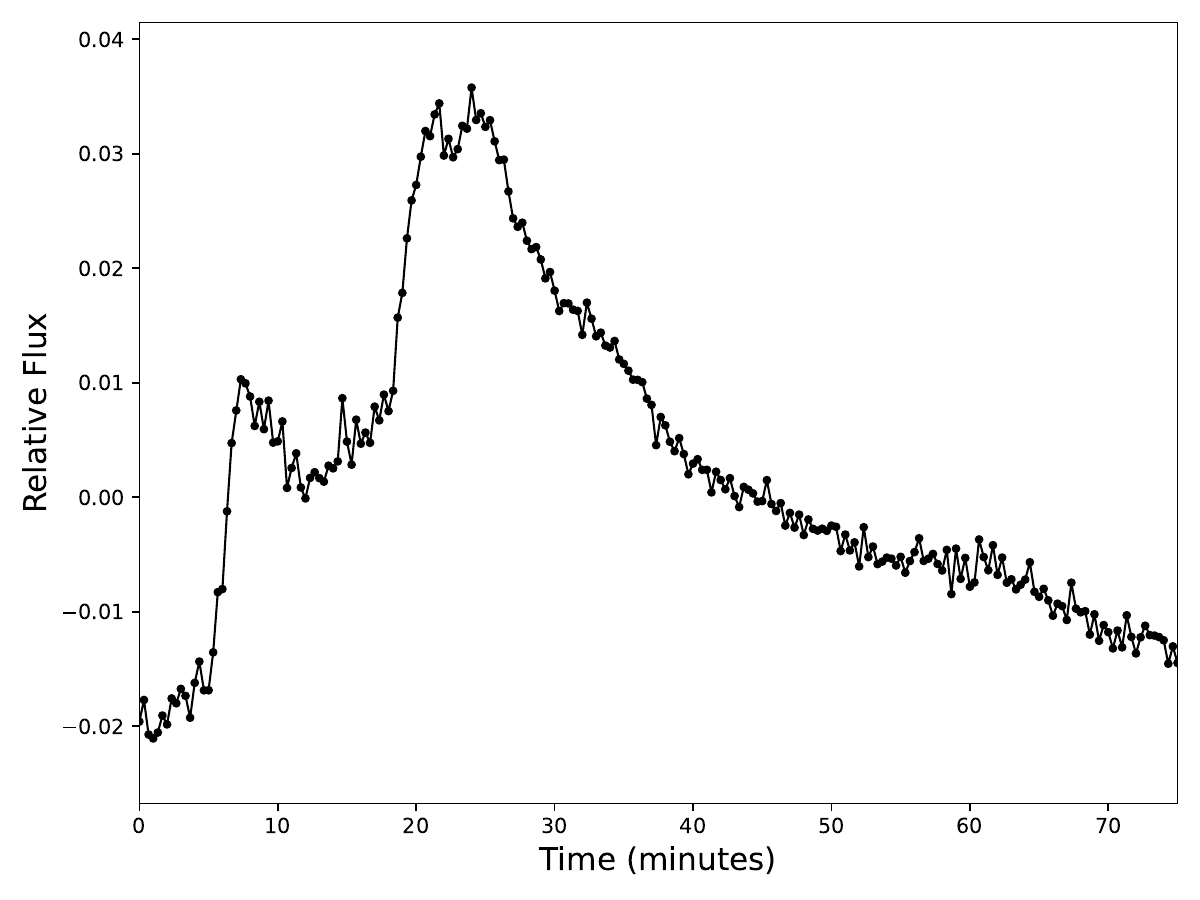}
\caption{Example of a flare with a flat-top morphology (left) and a complex multi-peak structure (right). Such flares are harder to detect in lower cadence observations such as the 2-minute TESS observations.}
\label{fig:complex_flares}
\end{figure*}

In Fig. \ref{fig:Energy_20s} we illustrate the flare-energy distribution and the power law fit. Our detection pipeline captures a relatively high number of lower-energy flares predominantly around 10$^{28}$ to 10$^{34}$ erg. The distribution peaks near 10$^{30}$ erg and significantly drops off beyond 10$^{34}$ erg. This has also been noted in \citet{2024A&A...689A.103Z}. We should also note that our detection method might also filter out slower flares. High-energy flares usually are longer lasting. ARIMA model is good at capturing gradual variations and can treat slowly varying signals as part of the baseline. If a flare rises and falls slowly, its shape might get “fit out” of the residual, making it fall below the 3$\sigma$ threshold.

We estimate the exponent $\alpha$ using maximum likelihood estimation for a continuous power-law model above a lower bound $X$ $\geq$ $X_{\text{min}}$, implemented via the powerlaw package in Python \footnote{https://pypi.org/project/powerlaw/}:
\begin{equation}
p(X) \;=\; \frac{\alpha - 1}{\,X_{\min}\,}
\left(\frac{X}{X_{\min}}\right)^{-\alpha}
\quad \text{for} \quad X \ge X_{\min}.
\end{equation}
Here, $p(X)$ denotes the probability density function of the variable $X$, which corresponds to the flare energy.
\noindent We calculate $\alpha$ by fitting the $\alpha$ above a certain threshold $X_\mathrm{min}$, in this case $X_{min}$ = 5.12 x 10$^{31}$ erg. We determine the $X_{min}$ by minimizing the Kolmogorov-Smirnov (KS) statistic, by comparing the empirical distribution to the theoretical power-law model. For solar flares, the typical power-law indices range for hard X-ray emissions is 1.5 to 1.7 \citep{2024LRSP...21....1K}. \citet{2021A&A...645A..42I} calculated the power-law for optical stellar flares using K2 observations and found the typical power-law ranges from 1.8 to 2.6. \citet{2019ApJS..241...29Y} calculated the power law indices for A, F, G, K and M stars separately and the $\alpha$ ranged from 1.12 to 2.13. They also found that the slope of the power law becomes steeper for K and M-type stars. \citet{2017ApJS..232...26V} analyzed the quarter 15 data from the Kepler mission and found that F-type stars exhibit a shallower slope than G-type and K–M dwarfs, while giant stars show a slope that falls between these two extremes. In our study, we found the value of $\alpha$ = 1.65. This reduced power-law slope associated with TESS’s red-optical sensitivity was similarly reported by \citet{2022ApJ...935..143P} who found $\alpha$ ranging from 1.5 to 1.7 depending on which flare energy calculation methodology is used. We note that the flare energies found in the TESS survey are generally higher than in Kepler. This is because TESS uses a red-optical bandpass filter, which is more sensitive to longer wavelengths of light compared to the filter used by Kepler \citep{2021ApJS..253...35T}.

In a similar way, we calculated the power-law slope for the ED as shown in Fig. \ref{fig:ED_20s}. Previous studies such as \citet{2019A&A...622A.133I, 2021A&A...645A..42I} which calculated the $\alpha$ of ED generally agree with our result of $\alpha$ = 2.62. This scaling exponent indicates that with each incremental increase in ED, the frequency of flares diminishes at a predictable rate dictated by the power-law. The ED in Fig. \ref{fig:ED_20s} displays a strong peak within a short ED range, particularly from around 10$^2$ to 10$^4$ seconds. A similar power-law fit is noted in Fig. 4 from \citet{2021A&A...645A..42I} for data from 30-minute cadence K2 mission.

Another advantage of observing in higher cadence is the fact that we can observe finer substructures in the flares \citep{2022ApJ...926..204H}. These observations enable the detection of intricate substructures and QPPs in flares that are often obscured by lower cadences. Consequently, instead of appearing as single, smooth brightness spikes, flares in high-cadence data display multiple peaks and genuine oscillatory behaviors which when modeled can also be used to detect the QPPs in them \citep{2016ApJ...829...23D}.  In \citet{2022ApJ...926..204H} 46$\%$ of flares exhibit substructures in the rise phase of the flare. They also notice the flare peaks have variations which may be due to rounded peak morphologies or slight misalignment when standardizing time axes using FWHM. 30$\%$ of all flaring events observed in the Characterizing Exoplanets Satellite (CHEOPS) and TESS flares in the work by \citet{2024A&A...686A.239B} had complex substructure. Some of these flares exhibited QPP-like signatures. In our analysis, we also notice such complex flares as shown in Fig. \ref{fig:complex_flares}.  We can observe that both the flares have complex substructures within them. While the flare in the right panel, which exhibits a more complex rise phase, has been observed in 46$\%$ of the flares in \citet{2022ApJ...926..204H}, it is thought that such rise‐phase complexity arises because multiple regions on the star’s surface become active. Alternatively, these multiple structures could also be interpreted as a QPP event. These regions might be adjacent or connected but are not necessarily ignited simultaneously. Such an event was observed during the 2000 Bastille Day solar flare \citep{2001ApJ...550L.105K} where two flares with an interconnected ribbon produced a light-curve with two peaks \citep{2019ApJ...871..167K}. In Fig. \ref{fig:complex_flares}, the left panel shows flares that have a flat-top morphology meaning the flares have a constant emission level at the peak. \citet{2021MNRAS.504.3246J} noticed such a flare in the Next Generation Transit Survey.
\section{Searching for Quasi-Periodic Pulsations}
\label{sec:qpps}
\begin{figure*}[h!]
\centering
\includegraphics[width=18cm]{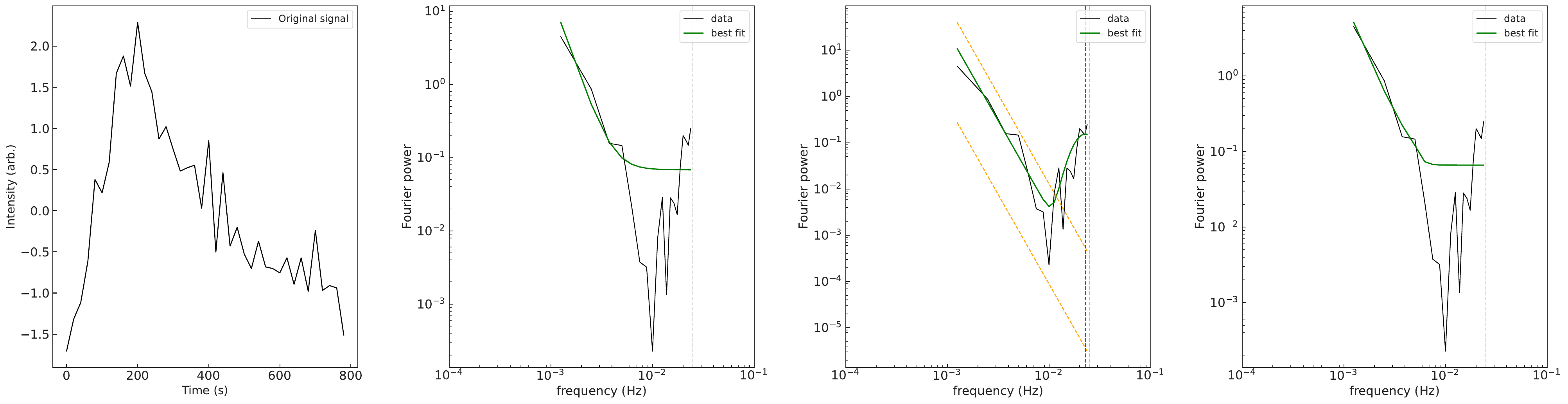}
\includegraphics[width=18cm]{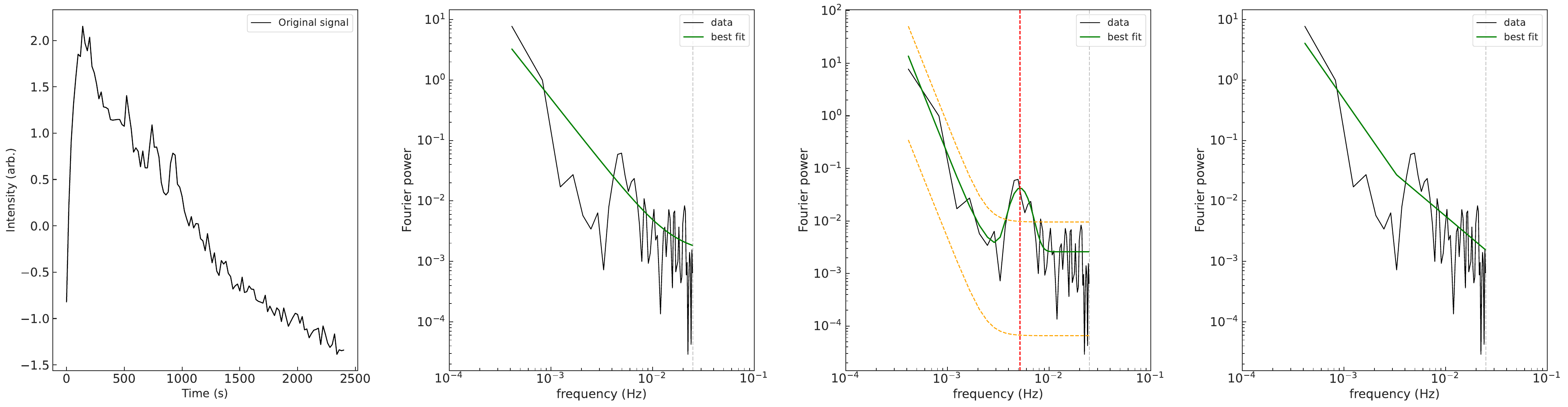}
\caption{Example of AFINO models applied to a stellar flare detected in TIC 33878971 (top panel) and TIC 303443210 (bottom panel). From left to right: 
(1) Flare light curve showing the observed flare variations, highlighting the detected QPP peaks. 
(2) Single power-law model fit to the power spectrum (solid line), highlighting no QPP detection. 
(3) Power-law model with a Gaussian bump (solid line), demonstrating excess power at a specific frequency indicative of quasi-periodic pulsations. The vertical dashed red line denotes the $f_{p}$, the location of the bump in the PSD. The dashed orange line indicates the $\pm$ 2$\sigma$ confidence interval around the fitted power-law component.
(4) Broken power-law model fit.}
\label{fig:afino_example}
\end{figure*}
\subsection{Employing AFINO for QPP detection}
\label{subsec:afino}
To search for QPPs in complex flare light curves, several methods have been be explored. For example, \citet{2016MNRAS.459.3659P} used wavelet transform to detect periodicities in Kepler flare light curves. \citet{2024ApJS..274...31B} developed a Fully Convolutional Networks based method that searches for exponentially decaying harmonic QPP. \citet{2021SoPh..296..162R} used a combination of methods such as Fourier transform with and without detrending as well as empirical mode decomposition. \citet{2015ApJ...798..108I} introduced a method based on examining the Fourier power spectrum and model fitting to find global QPP signatures in the flare time-series. This method was developed as flares display a power-law distribution in their Fourier spectra \citep{2011A&A...533A..61G, 2019ApJS..244...44B}. In this work, we utilize the Automated Flare Inference of Oscillations (AFINO) method presented by \citet{2015ApJ...798..108I}.

The choice to utilize AFINO is two-fold: Firstly, we want to test AFINO's efficacy for detecting QPPs in 20s cadence TESS light curves. In a preliminary analysis by the developers of AFINO, it was tested on the 2-minute cadence TESS light curves and it was able to recover 5 QPPs out of the 45 flares analyzed from 6 M-dwarfs. However, such an analysis was not carried out with the 20s cadence mode as it was not available during their testing. Secondly, in a blind test performed in \citet{2019ApJS..244...44B} with several other QPP detection methods, it was found that AFINO has the lowest false alarm probability. This becomes especially important in TESS detections, due to the fact that even in 20s observations, the total number of data points remains limited when compared with the high-cadence observations typically employed in solar flare studies. This scarcity of data points increases the susceptibility to statistical fluctuations and noise, which can masquerade as false QPP signals. However, that also means AFINO is particularly conservative and has one of the lowest detection rates of all the methods it was compared with \citep{2019ApJS..244...44B}.

AFINO works by examining the Fourier power spectrum of the flare and performing a model comparison by fitting three different models. Before fitting the three models, it normalizes the time-series data by the mean and applies a Hann-window function. Then it computes the Fourier power spectrum and fits the three models: Model M0 is a single power law model, M1 which is a power law with a Gaussian bump, and finally model M2 is a broken power-law model that adds flexibility to fit complex noise structures. Model M2 is designed with the idea that a single power-law model may not be enough to fit the flare Fourier power spectrum. Model M1 which is the QPP model has the additional Gaussian component in the log-log frequency space. Since the goal of our analysis is to detect short-period QPPs, we restricted the frequency $f_{p}$ to be $\leq$ 300 s. As the TESS data is binned at 20 s cadence, AFINO cannot detect pulsations below the Nyquist frequency of 0.025 Hz or equivalently 40 s. To identify the presence of a QPP, AFINO evaluates the goodness-of-fit for each model using a maximum likelihood estimator. A preferred model \( M_i \) must satisfy the condition:
\begin{equation}
\label{eqn:model_criterion}
\Delta \text{BIC} = \text{BIC}_{j} - \text{BIC}_i > 10 \quad \forall \, j \neq i,
\end{equation}
indicating strong evidence for model \( M_i \). We further validate these findings via the reduced $\chi^{2}$ statistic to confirm consistency between the model and the observed data. For a more thorough explanation of the method, we refer the reader to \citet{2015ApJ...798..108I, 2016ApJ...833..284I, 2019ApJS..244...44B}.

When comparing the results from AFINO with other works, we observed that some flare-associated pulsations extend beyond the initially detected flare time window after the end of the flare. These extended pulsations may not be captured within the standard flare duration, leading to an incomplete analysis of the flare's characteristics. To address this issue and ensure a comprehensive analysis of the flare events, we implemented a method to extend the flare time window. Specifically, we extended the end time of the flare by adding multiples of the flare's half duration $(\tau)$. It is calculated based on the flare's duration, defined as half the difference between the flare's start and end times:
\begin{equation}
\label{eq:sigma_definition}
\tau = \frac{\Delta t}{2} = \frac{t_{\text{end}} - t_{\text{start}}}{2}.
\end{equation}
We considered extensions of \( 1\tau \), \( 2\tau \), and \( 3\tau \) to the original flare end time:
\begin{equation}
\label{eq:extended_time}
t_{\text{extended}} = t_{\text{end}} + n \tau,
\end{equation}
This approach allows us to include additional data points that may contain significant signals occurring immediately after the initially detected flare duration.

\begin{figure*}[htbp!]
\centering
\begin{minipage}{0.48\textwidth}
\includegraphics[width=\linewidth]{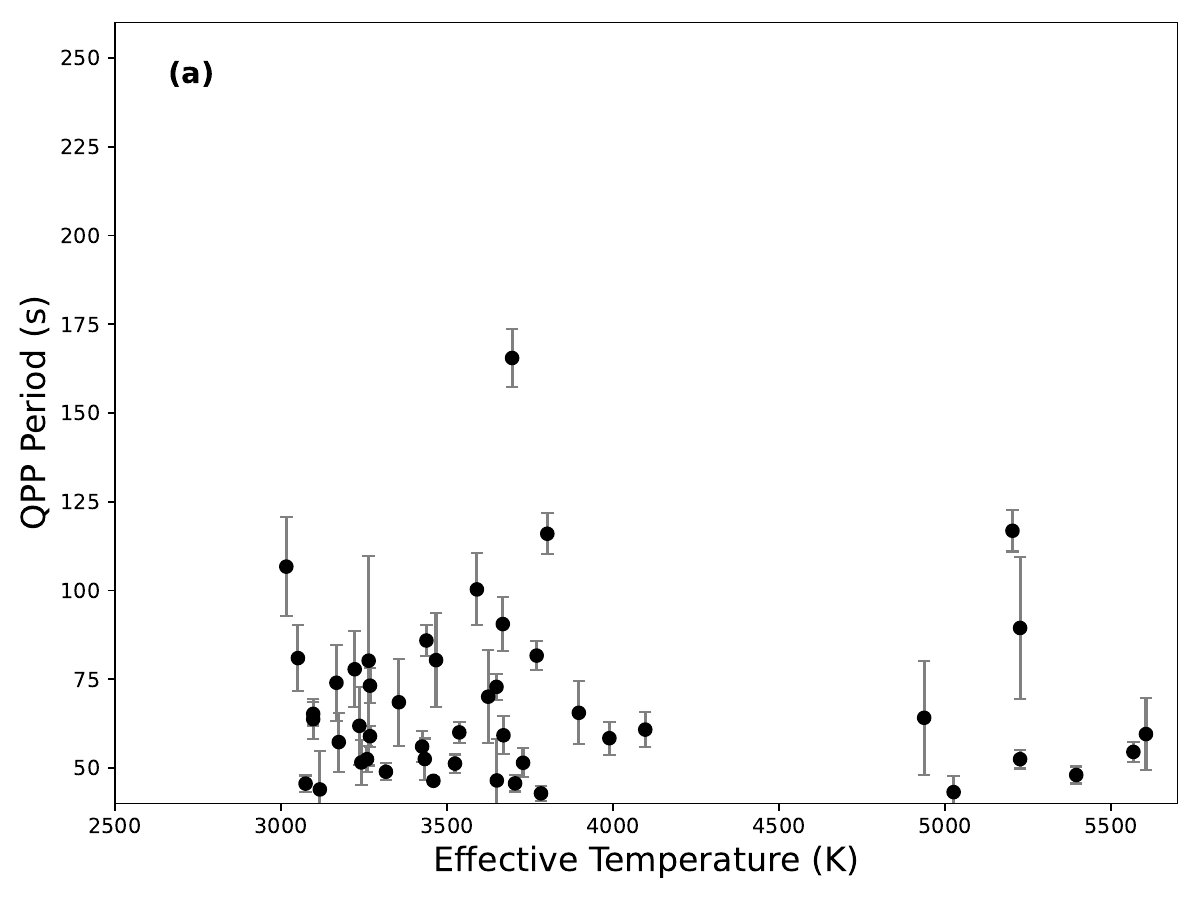}
\end{minipage}
\hfill
\begin{minipage}{0.48\textwidth}
\includegraphics[width=\linewidth]{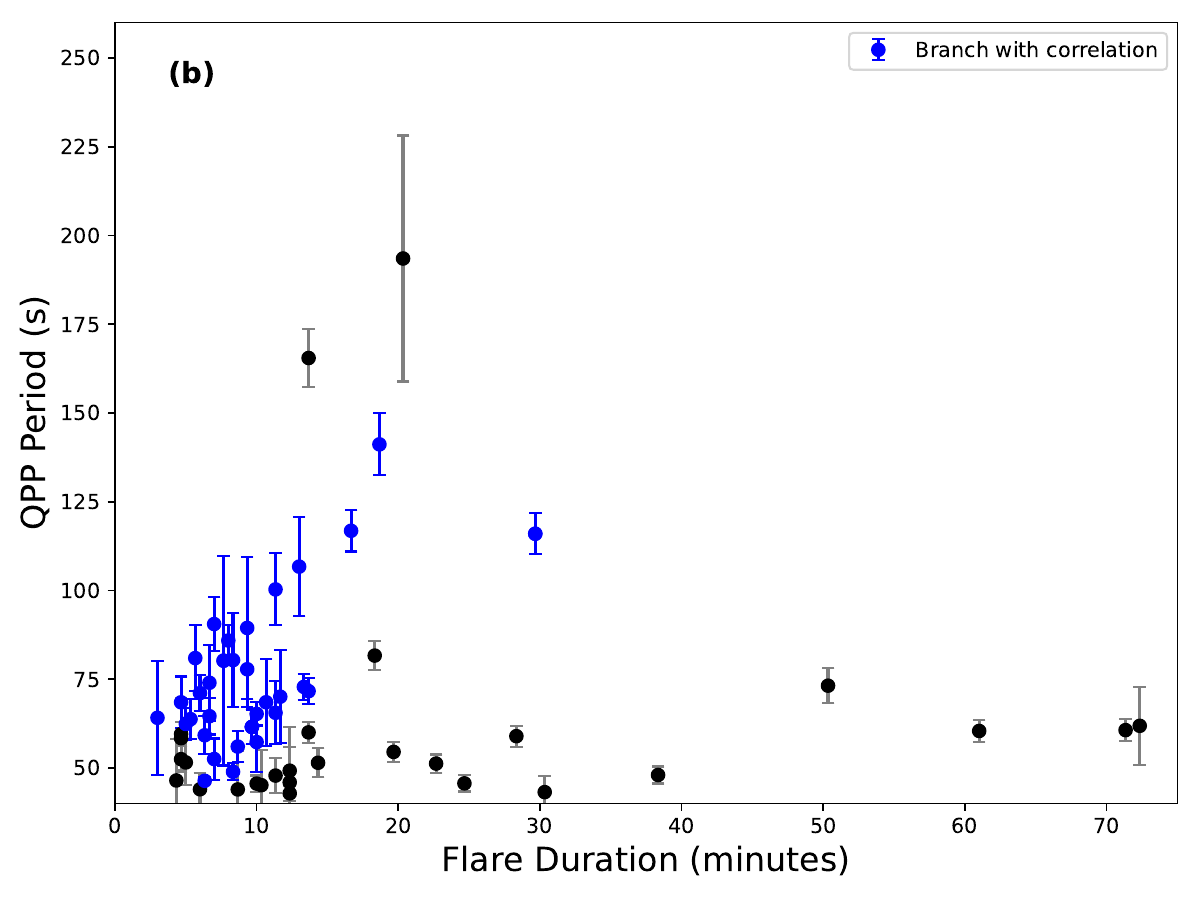}
\end{minipage}
\vspace{1em}
\begin{minipage}{0.48\textwidth}
\includegraphics[width=\linewidth]{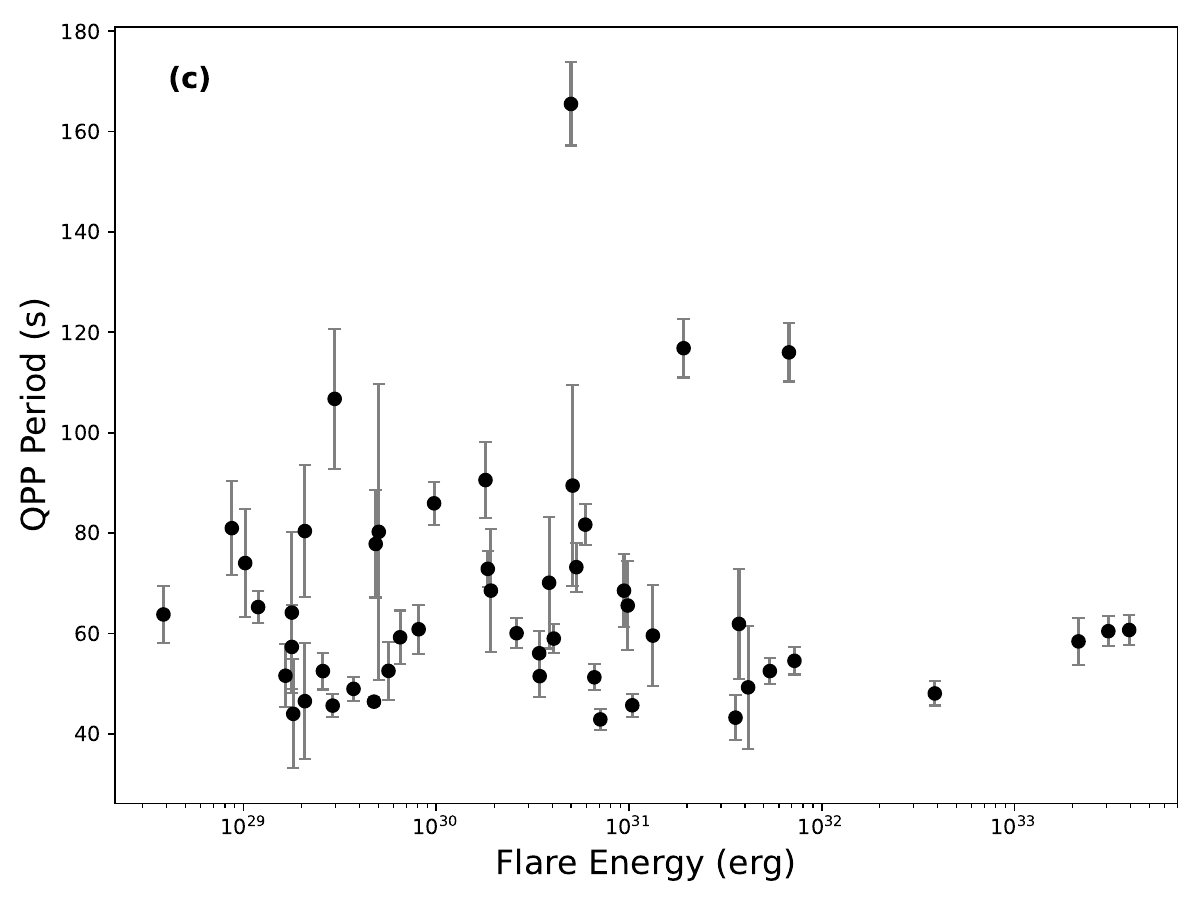}
\end{minipage}
\hfill
\begin{minipage}{0.48\textwidth}
\includegraphics[width=\linewidth]{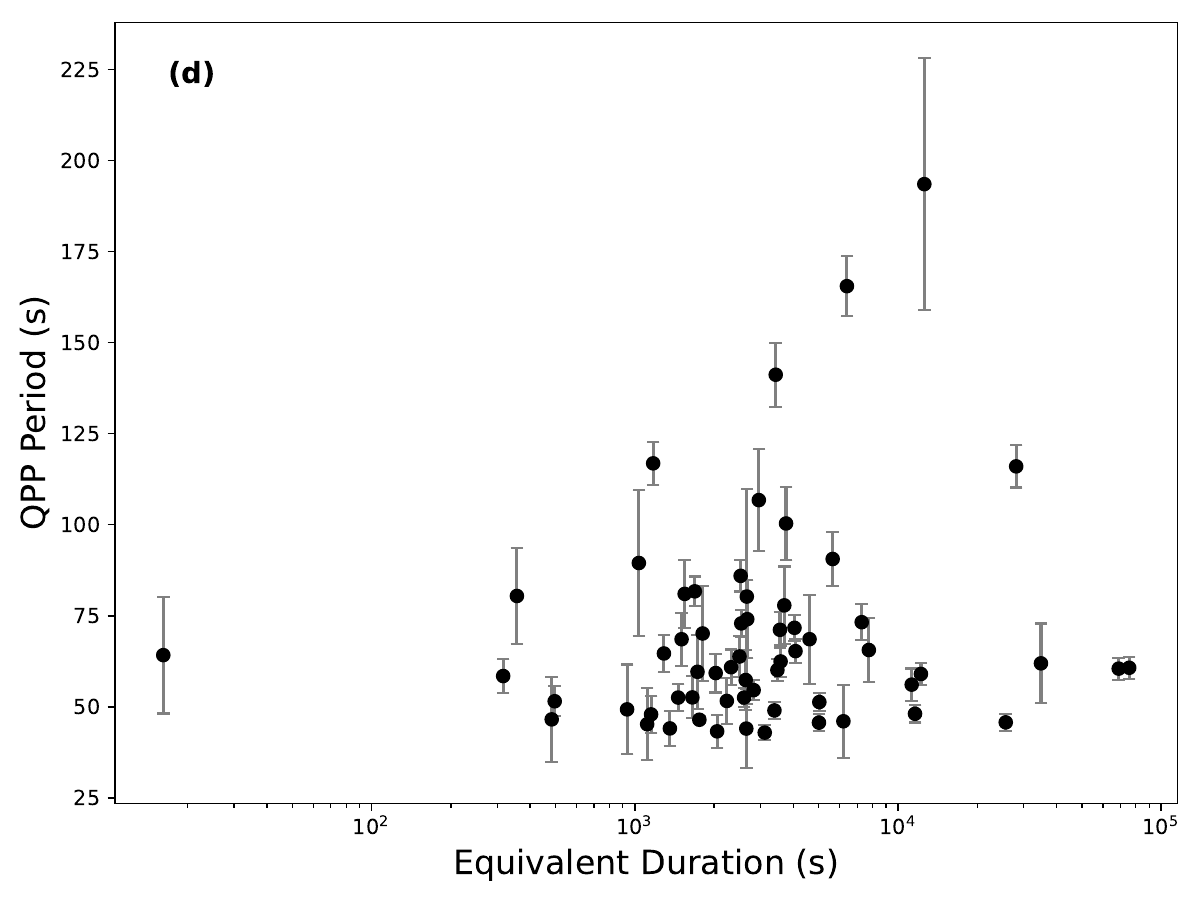}
\end{minipage}
\vspace{1em}
\begin{minipage}{0.48\textwidth}
\includegraphics[width=\linewidth]{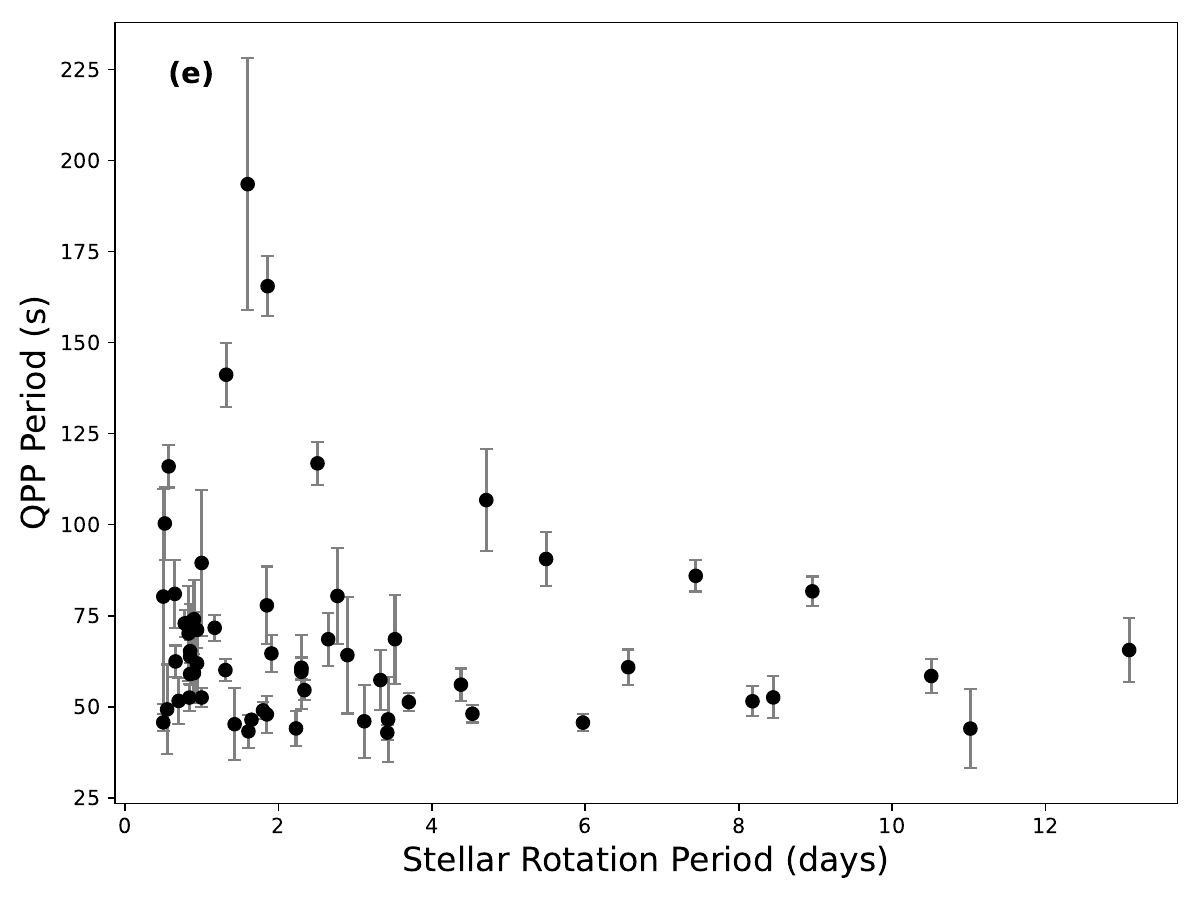}
\end{minipage}
\hfill
\caption{QPP periods with associated errors against various parameters: (a) Stellar effective temperature (b) Flare duration along with the branch indicating a correlation highlighted in blue (c) Flare energy (d) Equivalent duration (e) Rotation period of the star.}
\label{fig:stellarparams}
\end{figure*}
\subsection{Results from AFINO analysis}
From our analysis of 3878 flares, we found 61 QPP-like signatures in 57 individual stars. The period of the detected QPPs ranges from 42 seconds to 193 seconds. When we compare our findings to the work by \citet{2016MNRAS.459.3659P, 2021SoPh..296..162R, 2021csss.confE.272M, 2022ApJ...926..204H}  which search for QPPs in Kepler and TESS, it is immediately clear that AFINO excels at detecting short period QPPs. In the previous studies, the period of QPPs are on the order of a few minutes to tens of minutes. In a recent survey for the Kepler catalog, \citet{2024ApJS..274...31B} found 7$\%$ of their detected events had QPP-like signatures. In \citet{2022ApJ...926..204H} found 17 confirmed QPPs in 3792 flares. \citet{2016MNRAS.459.3659P} found 2.5$\%$ of events had QPPs in their Kepler survey. On the other hand, in our study, we find 1.6$\%$ of flare events have QPP-like signatures.

As discussed in Sect. \ref{subsec:afino}, AFINO’s conservative nature stems from its assumption of weakly decaying oscillations (particularly in model M1) which leads to a lower detection rate relative to other methods. This assumption favors the detection of oscillations that maintain a relatively constant amplitude over time. Short-period oscillations are more likely to meet this criterion within the typical duration of flare observations. Conversely, longer-period QPPs that decay quickly might not significantly enhance the PSD. This is because rapid decreases in oscillation amplitude lead to diminished signal power, reducing its presence in the PSD. Fewer oscillation cycles are completed within the observation window, making it harder to resolve the period accurately. It is also worth noting, AFINO cannot detect multi-period oscillations as it is primarily designed to detect the presence of single periodicities in flare time series data. AFINO assumes that the oscillatory signals are stationary over the time interval analyzed, meaning their properties do not change with time. Multi-period oscillations include signals that vary in amplitude or phase or that switch dominance over time. However, AFINO is made for stationary periods only. For QPPs with multi-periodicity, the wavelet approach as used by \citet{2016MNRAS.459.3659P} will be more suitable. In contrast, when \citet{2024ApJS..274...31B} applied their QPP detection method to the AFINO catalog of detections from 30 AFINO QPPs, their method was only able to find one oscillation. This also indicates how different methods can detect different types of QPPs.

Table \ref{table:qpp_table} shows our QPP detections along with their properties. Five stars in our catalog have two QPP detections each: TIC 220433364, 348898049, 303443210, 429354344, 345451496. Three are M-dwarfs, one is K type and one is a B-type subdwarf \citep{1994ApJ...432..351S}. Although the periods of the pulsations for the M and K-type stars are largely different even for QPPs on the same star, the periods detected on TIC 345451496 which is a B-type subdwarf for both the flares are $\sim$ 60s. This repeated timescale is especially intriguing given that hot subdwarfs lack the kind of deep convective envelopes typically associated with magnetically driven flaring activity \citep{2012A&A...541A.100L}. One may wonder if these short, $\sim$ 60 s pulsations truly originate within the thin, hot envelope of the subdwarf itself or if a hidden companion could instead drive them. Given that the period remains consistent across two separate flares, such a stable mechanism merits further investigation. 

In Fig. \ref{fig:afino_example} we show examples of QPPs detected by AFINO in the flare lightcurve of TIC 33878971 and TIC 303443210. The top panel shows a QPP with a period of ~44s and the bottom panel shows a QPP that has a period of 193.5s which is the highest period QPP we found. We also note that in some previous AFINO studies carried out on solar flares, the flaring lightcurve was broken into the impulsive and the decay phase of the flares and analyzed individually \citep{2019ApJ...875...33H, 2024ApJ...971...29I}. However, due to the 20s cadence of TESS, such an approach would yield only a small number of data points in each sub-interval. Thus, in the present work, we have analyzed the complete flare interval (from start to end) as a single segment, consistent with the original methodology described by \citet{2015ApJ...798..108I}.
\begin{table*}
\centering
\caption{QPP events detected in the dataset.}
\setlength{\tabcolsep}{8pt}
\resizebox{\textwidth}{!}{
\begin{tabular}{lccccccc}
\hline\hline
\noalign{\smallskip}
TIC ID & Start Time (TBJD) & End Time (TBJD) & Period (s) & Error P (s) & $T_{\rm eff}$ (K) & $P_{\mathrm{rot}}$ (d) & \dots \\
\noalign{\smallskip}
\hline
\noalign{\smallskip}
231914259 & 2071.284 & 2071.302 & 45.19274 & 9.942 & -- & 1.43 & \dots \\
355766445 & 2070.211 & 2070.227 & 51.26799 & 2.563 & 3524 & 3.7 & \dots \\
358176584 & 2067.428 & 2067.434 & 59.23968 & 5.331 & 3670 & 0.9 & \dots \\
\noalign{\smallskip}
\hline
\multicolumn{8}{c}{\dots \dots \dots} \\
\hline
\noalign{\smallskip}
233342788 & 3377.944 & 3377.969 & 45.694 & 2.284 & 3705 & 0.5 & \dots \\
11333775 & 3455.940 & 3455.950 & 61.59393 & 4.927 & 4097 & 6.56 & \dots \\
350166984 & 3493.082 & 3493.085 & 64.16237 & 0.25 & 4937 & 2.9 & \dots \\
\noalign{\smallskip}
\hline
\end{tabular}}
\tablefoot{Columns include: TIC ID, start and end time of the flare, period of the QPP in seconds, period error, effective temperature in (K), and stellar rotation period in days. The complete table includes additional parameters such as tau, time extension, best model BIC, power-law index, reduced $\chi^2$ values for each model, flare energy, duration, and amplitude. The full table can be found at the GitHub
repository: https://github.com/aadishj19/QPPs-in-TESS-flares}
\label{table:qpp_table}
\end{table*}
\begin{figure}[h!]
  \resizebox{\hsize}{!}{\includegraphics{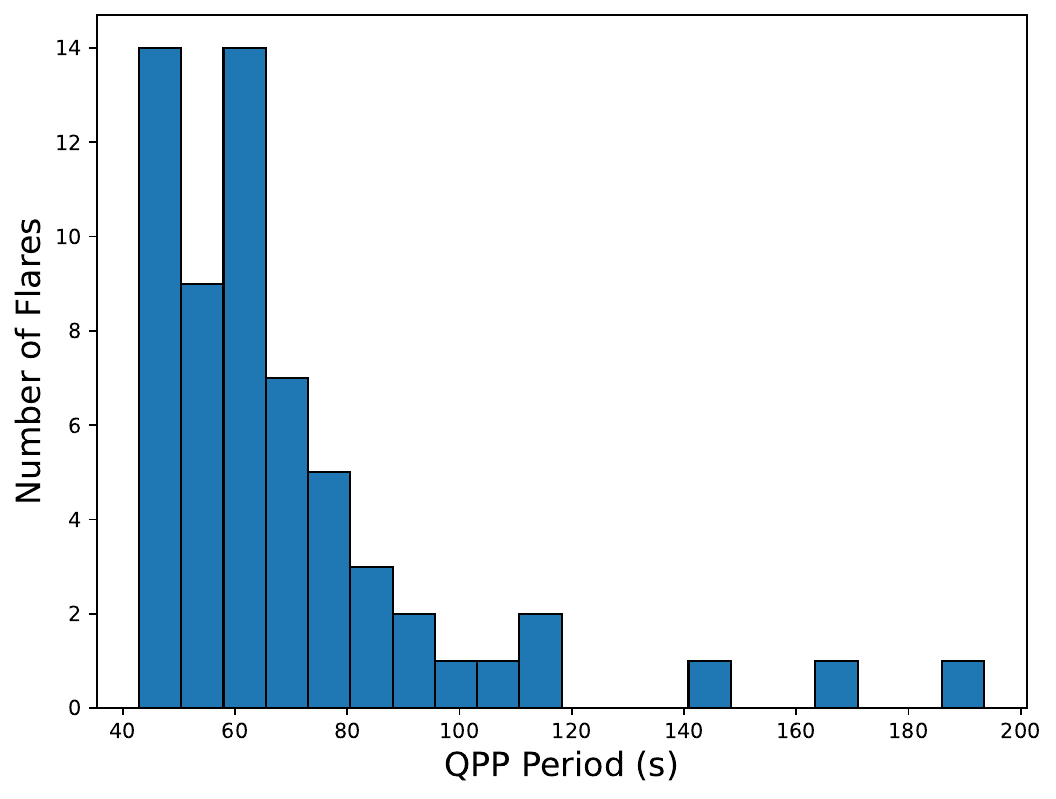}}
  \caption{Histogram of QPP periods detected in our sample. The distribution reveals a median period of 60.8 seconds.}
  \label{fig:qpp_hist}
\end{figure}
In Fig. \ref{fig:stellarparams} we show the QPP periods against various stellar parameters. To check if there is a linear correlation between the oscillation period and either the stellar or flare parameters, we performed the Pearson correlation test. Consistent with prior findings, our results indicate that QPP periods do not appear to depend on either global stellar parameters (effective temperature, rotation rate) or on flare-specific parameters (energy, ED). The Pearson correlation tests show that the period bears no relationship with effective temperature (r = -0.041, p = 0.777), flare duration (r = -0.042, p = 0.767), flare energy (r = 0.080, p = 0.576), equivalent duration (r = -0.010, p = 0.941), or the stellar rotation period (r = -0.136, p = 0.292). A similar result has been found in other studies such as \citet{2016MNRAS.459.3659P}, which focused on QPPs of a few minutes’ duration. Here, we confirm that this relationship (or lack thereof) also holds for the shorter QPP periods examined in our study. Fig. \ref{fig:qpp_hist} we plot the histogram of the QPPs periods. It is clear that most of the QPPs have periods below 80 seconds. When compared to the AFINO results of \citet{2016ApJ...833..284I} for solar flare QPPs observed in GOES X-ray data, they reported a median period of 17.5 seconds.
\begin{figure*}
\centering
\includegraphics[width=0.48\textwidth]{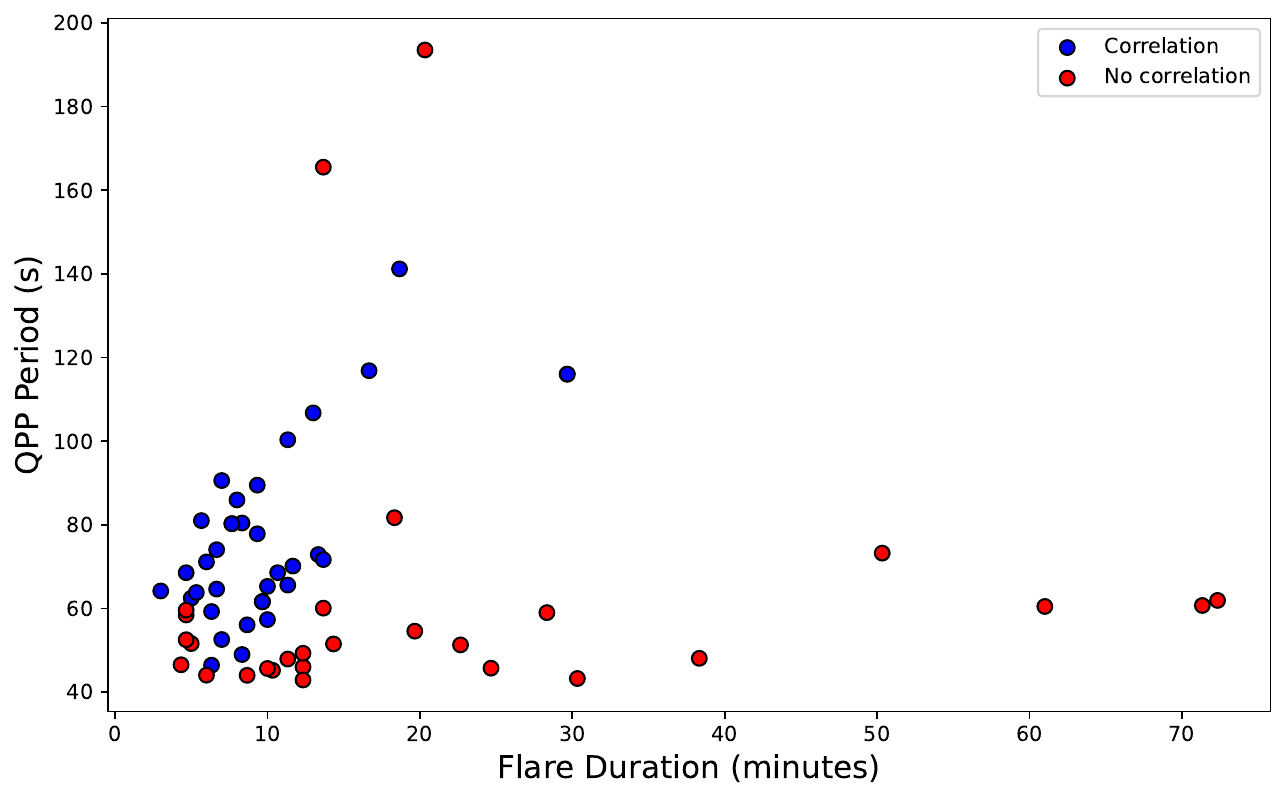}
\hfill
\includegraphics[width=0.48\textwidth]{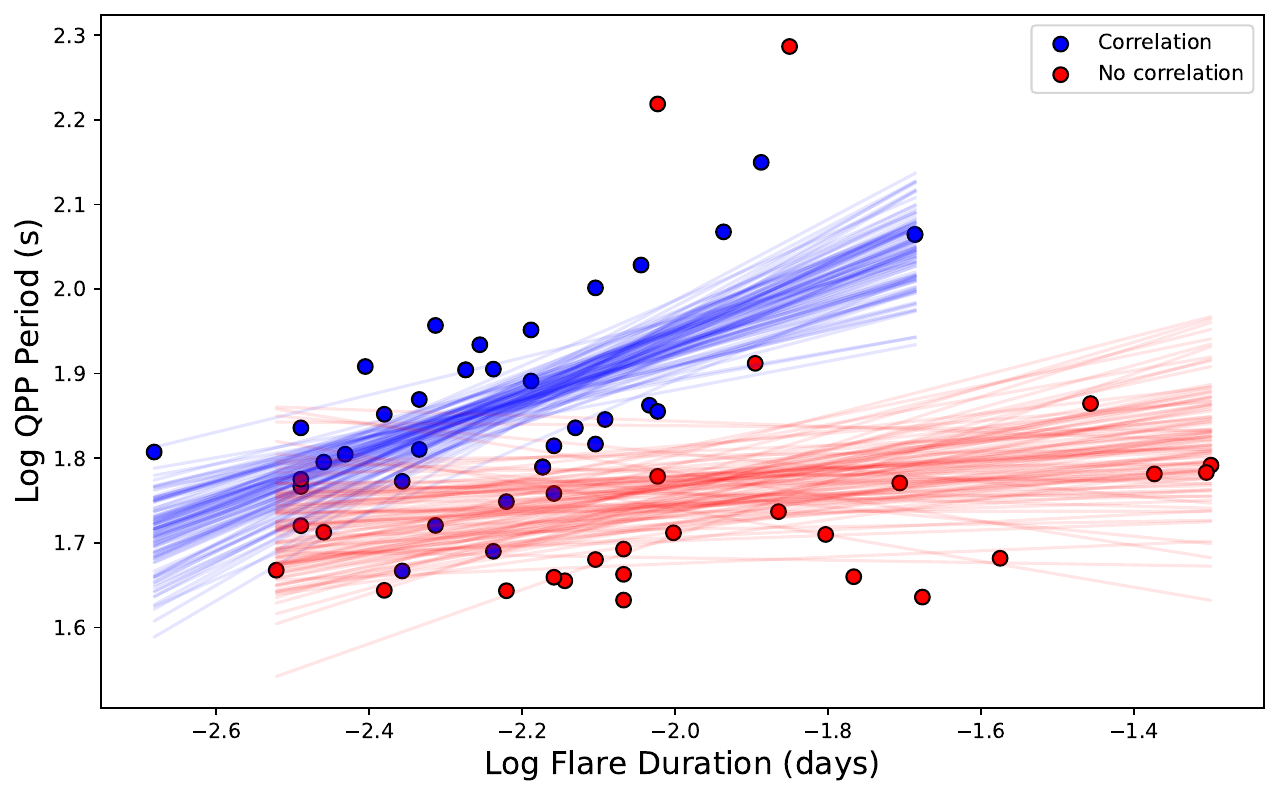}
\caption{QPP periods against flare duration showing a visual correlation in blue (left) with the Bayesian linear fit applied to the same dataset (right).}
\label{fig:fit}
\end{figure*}
\begin{figure*}
\centering
\includegraphics[width=0.48\textwidth]{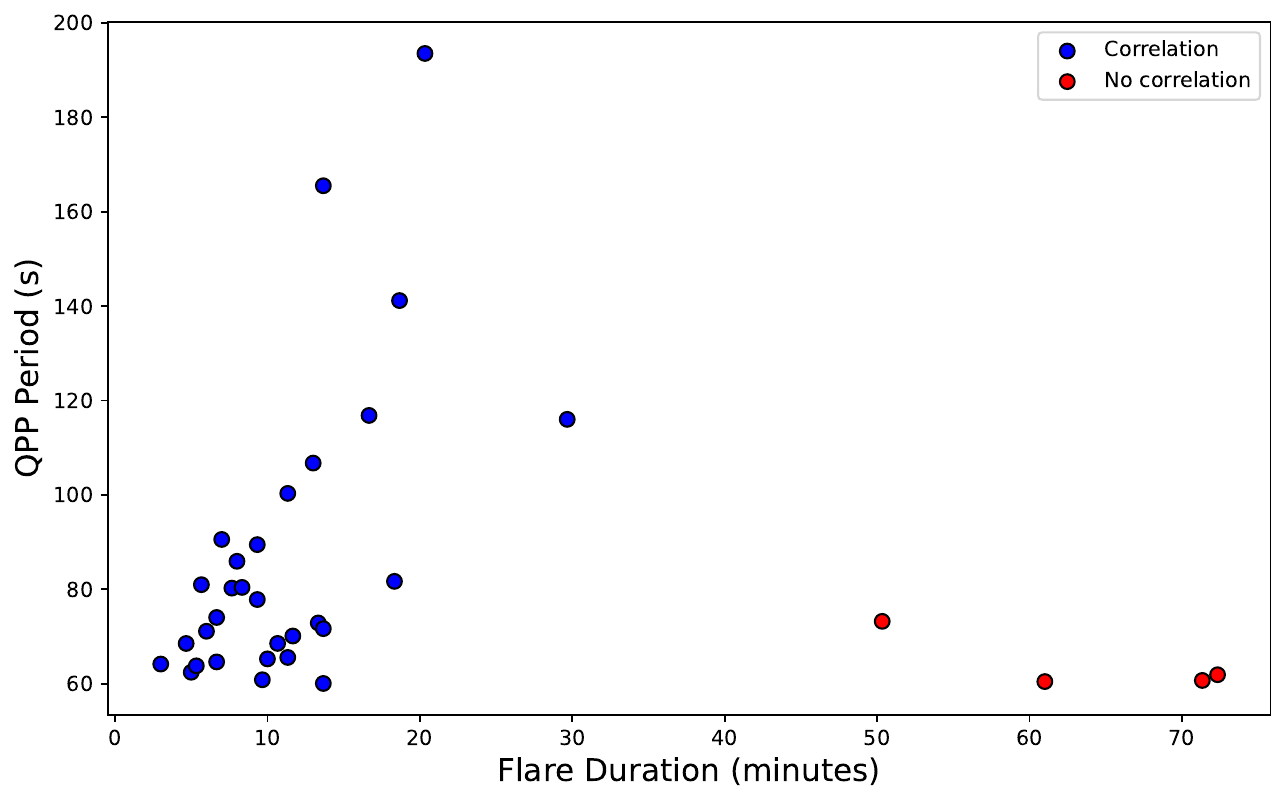}
\hfill
\includegraphics[width=0.48\textwidth]{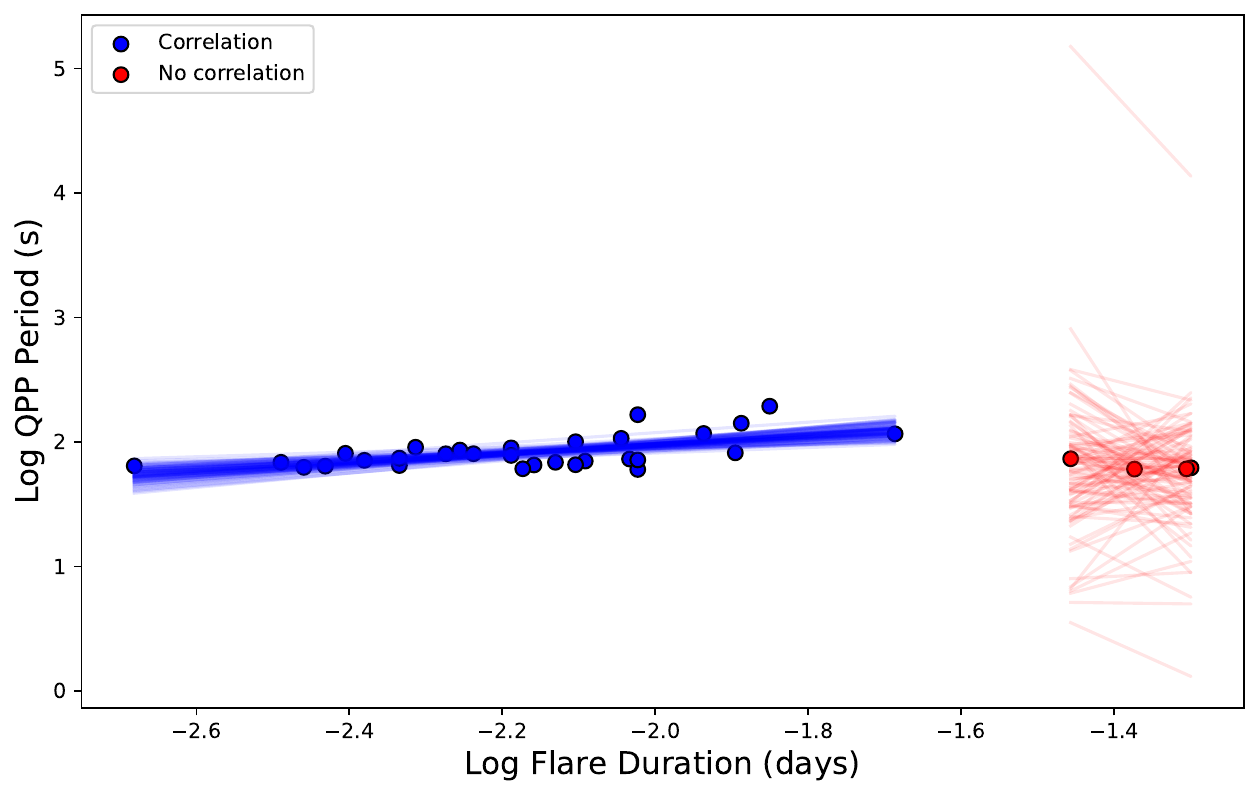}
\caption{QPP periods above 60 seconds against flare duration showing a visual correlation in blue (left) with the Bayesian linear fit (right).}
\label{fig:fit_new}
\end{figure*}
On visual inspection, we find a branch of QPPs showing a linear correlation as highlighted in Fig. \ref{fig:stellarparams} (b) and in Fig \ref{fig:fit}. To separate the branch of QPPs for the analysis from the rest of the data points, K-means clustering was applied. We find that the optimal number of clusters is $K$ = 4, where clusters 1 and 2 included the branch showing a visual correlation and clusters 3 and 4 included the rest of the data points. As clusters 1 and 2 indicated a similar slope, we grouped them together as a single cluster for analysis. Similarly, we grouped together cluster 3 and 4. A similar correlation was also observed in the GOES X-ray data for solar flares, indicating that longer-duration flares tend to exhibit QPPs with longer periods \citep{2017A&A...608A.101P, 2019A&A...624A..65P, 2020ApJ...895...50H}. In order to determine if our dataset also contains a positive correlation, we took a similar approach to \citet{2020ApJ...895...50H}. We first performed both Pearson and Spearman correlation tests to detect any potential linear or monotonic trends. Then, we applied a Bayesian linear regression in log–log space for the QPP branch visible in our scatter plot. We find that there is a statistically significant positive correlation between the QPP period and flare duration for the branch that showed a visual correlation. The Spearman rank correlation coefficient is 0.42 (p-value: 0.0112), indicating a strong monotonic trend. Additionally, the Pearson correlation coefficient is 0.69 (p-value: 0.0000), suggesting a strong linear relationship in log-log space. Our analysis yields the following scaling equation for the relationship between QPP periods (\(P\)) and flare durations (\(\tau_{\text{flare}}\)):
\begin{equation}
\log P = (0.33 \pm 0.08) \log \tau_{\text{flare}} - (2.59 \pm 0.17)
\end{equation}
This implies a power-law relationship expressed as:
\begin{equation}
P \propto \tau_{\text{flare}}^{(0.33 \pm 0.08)}
\end{equation}
As for the branch which did not indicate any correlation visually, we performed the same steps as above. We find that the Spearman rank correlation coefficient is 0.36 (p-value: 0.0581) suggesting that the relationship is not statistically significant. Meanwhile, Pearson’s correlation coefficient is 0.05 (p = 0.7943), indicating negligible linear correlation. Overall, these findings do not support the presence of a meaningful relationship. We obtain the following relationship:
\begin{equation}
\log P = (0.08 \pm 0.09) \log \tau_{\text{flare}} - (1.91 \pm 0.18)
\end{equation}
For the branch indicating a correlation, the exponent value of \(0.33 \pm 0.08\) suggests that the increases in flare duration are associated with increasing QPP periods, at a sublinear rate. This finding is consistent to that identified in previous studies of solar flares \citep{2017A&A...608A.101P, 2019A&A...624A..65P, 2020ApJ...895...50H}. Additionally, we investigated whether this correlation might be influenced by the stellar rotation rate or by the measurement uncertainties in the QPP periods obtained via AFINO. We found no significant relationship with either quantity, indicating that the observed correlation between the QPP period and flare duration is robust against these factors. We further examined whether the observed correlation persists after excluding QPP events with periods shorter than 60 seconds i.e., less than three times the sampling interval (3 × 20). This threshold was imposed to eliminate signals near the Nyquist limit and ensure that the analysis relies only on well‐resolved periodic features. We found that we are left with 34 QPPs as shown in Fig. \ref{fig:fit_new}. After applying similar steps as the original dataset however this time with K=2 clustering, we found the Spearman and Pearson correlation coefficient of 0.52 (p-value: 0.0034) and 0.60 (p-value:0.0005) respectively for the branch showing a correlation. This results in the following scaling relationship:
\begin{equation}
\log P = (0.34 \pm 0.10) \log \tau_{\text{flare}} + (2.67 \pm 0.21)
\end{equation}
And the power-law relationship:
\begin{equation}
P \propto \tau_{\text{flare}}^{(0.34 \pm 0.10)}
\end{equation}
It is interesting to note that after removing the data-points below 60 s, most of the data-points show a linear correlation, unlike Fig. \ref{fig:fit}.  Moreover, the recent work by \citet{2024arXiv241207580P} reinforces our findings by demonstrating a similar correlation in their sample of 13 M-dwarf flares in the U-band, supporting the notion that longer-duration flares are generally associated with longer QPP periods. Interestingly, Fig. 7 in the work by \citet{2024arXiv241207580P} reveals striking parallels to our period–duration correlation. Both studies focus on flares of roughly 30 minutes in duration, where a discernible positive trend between the QPP period and flare duration emerges. 

Incorporating these results into the broader body of QPP research suggests that stellar flare processes may follow analogous scaling laws to those observed in the solar context, as previously determined by \citet{2017A&A...608A.101P, 2019A&A...624A..65P, 2020ApJ...895...50H}. Moreover, the slope of the power-law fit from our analysis (0.33) is similar to the slope reported by \citet{2020ApJ...895...50H} for solar flares (0.67), reinforcing the notion that similar scaling laws govern both stellar and solar flare processes.

\section{Conclusions}
\label{sec:conclusion} 
In our analysis, we detected 3878 flaring events across 1285 stars. Across these flares, 61 QPP events were detected on 57 individual stars, making our study the largest TESS-based QPP catalog to date.

Our ARIMA-based flare detection routine reliably identifies genuine flares when applied to the 20s cadence light curves. Nevertheless, some limitations arise from the selected parameter grid, which may exclude optimal (p, d, q) combinations needed to fully capture stellar variability without compromising flare detection. Limiting $p$ and $q$ to values between 0 and 3 could further miss models that better model the baseline flux yet preserve flare signatures. Additionally, ARIMA models with $d=0$ assume stationarity after differencing, which may not hold for every stellar time series. Despite these challenges, the fairly conservative nature of our approach ensures that most of the detected flares are true flare events, enhancing confidence in our final flare sample.

We also demonstrated the efficacy of AFINO in detecting QPPs in TESS data. The periods of these QPPs ranged from 42 to 193 seconds. Statistical tests indicated no significant correlations between QPP periods and effective temperature, flare energy, rotation period, and ED. We did find a positive correlation between QPP periods and the duration of the flare for a branch which indicates larger flares host longer QPPs.  Our study also extends the catalog of short-period stellar QPPs. Other QPP detection methods rarely detect sub-minute QPPs in stellar flares whereas AFINO excels at it. The periods of the QPPs detected in our work also supports the idea that if MHD oscillations are the cause behind QPPs then seismologically probing the flare environment remains a possibility.

Our study indicates a scaling relationship between QPP periods and the flare duration, but it does so only for a branch of QPPs rather than for all the data points. Hence, we should exercise caution when interpreting this scaling relationship. Notably, when the analysis is restricted to QPPs with periods exceeding 60 seconds, the scaling relationship becomes much more pronounced. In order to confirm if such a relationship does exist in stellar flares, more studies are required.  Future studies could include QPP detection made using a combination of methods that can cover the entire period range similar to solar flares so a single comprehensive catalog can be obtained. QPPs detected in this study are all detected in white light. White-light flares are still not well understood and the mechanism behind their generation remains a mystery. Our result serves as an important step toward confirming and generalizing these findings across larger samples.
\section*{Data availability}
Tables of the flare and QPP detections are available at: https://github.com/aadishj19/QPPs-in-TESS-flares
\begin{acknowledgements}
      The research that led to these results was subsidised by the Belgian Federal Science Policy Office through the contract B2/223/P1/CLOSE-UP. DL was supported by a Senior Research Project (G088021N) of the FWO Vlaanderen. TVD was supported by the C1 grant TRACEspace of Internal Funds KU Leuven and a Senior Research Project (G088021N) of the FWO Vlaanderen. Furthermore, TVD and DJF received financial support from the Flemish Government under the long-term structural Methusalem funding program, project SOUL: Stellar evolution in full glory, grant METH/24/012 at KU Leuven. The paper is also part of the DynaSun project and has thus received funding under the Horizon Europe programme of the European Union under grant agreement (no. 101131534). Views and opinions expressed are however those of the author(s) only and do not necessarily reflect those of the European Union and therefore the European Union cannot be held responsible for them.\\
      
      The authors express gratitude to the TESS consortium for providing excellent observational data. This research utilizes publicly available data from the TESS mission, accessed via the Mikulski Archive for Space Telescopes. This research made use of the SIMBAD database, operated at CDS, Strasbourg, France; the VizieR catalogue access tool, CDS, Strasbourg, France. \\

      \textit{Software}: AFINO \citep{2015ApJ...798..108I, 2016ApJ...833..284I}, Astropy \citep{2013A&A...558A..33A}, Lightkurve \citep{2018ascl.soft12013L}, pmdarima \citep{pmdarima}, SciPy \citep{2020SciPy-NMeth}, Pandas \citep{mckinney-proc-scipy-2010}, Matplotlib \citep{Hunter:2007}, Numpy \citep{harris2020array}, PyMC3 \citep{Salvatier2016}. 
\end{acknowledgements}
\bibliographystyle{aa}
\bibliography{bibliography/Literature.bib}

\end{document}